\documentclass[twocolumn,twocolappendix]{aastex631}

\usepackage{amsmath, amssymb}
\usepackage{graphicx,color}
\usepackage{float}
\usepackage{tabularx} %,siunitx}
\usepackage[caption=false]{subfig}
\usepackage{hyperref}

\begin{document}

\title{Constraints on the hosts of UHECR accelerators}

\author[0000-0003-4615-5529]{Marco Stein Muzio}
\email{msm6428@psu.edu}
\affiliation{Center for Cosmology and Particle Physics, Department of Physics, New York University, 726 Broadway, New York, New York, USA}
\affiliation{Department of Astronomy and Astrophysics, Pennsylvania State University, University Park, PA 16802, USA}
\affiliation{Department of Physics, Pennsylvania State University, University Park, PA 16802, USA}
\affiliation{Institute of Gravitation and the Cosmos, Center for Multi-Messenger Astrophysics, Pennsylvania State University, University Park,
PA 16802, USA}

\author[0000-0003-2417-5975]{Glennys R. Farrar}
\email{gf25@nyu.edu}
\affiliation{Center for Cosmology and Particle Physics, Department of Physics, New York University, 726 Broadway, New York, New York, USA}

\date{\today}

\begin{abstract}
Interactions of ultrahigh energy cosmic rays in the surroundings of their accelerators can naturally explain the observed spectrum and composition of UHECRs, including the abundance of protons below the ankle. Here we show that astrophysical properties of the UHECR source environment such as the temperature, size, and magnetic field can be constrained by UHECR and neutrino data. Applying this to candidate sources with a simple structure shows that starburst galaxies are consistent with these constraints, but galaxy clusters are in tension with them. For multi-component systems like AGNs and GRBs the results are indicative but customized analysis is needed for definitive conclusions.
\end{abstract}	

%\keywords{} 

\section{Introduction}

\par
The origin of ultrahigh energy cosmic rays (UHECRs with $E \gtrsim 10^{18}$~eV $= 1$~EeV), is a long-standing mystery. Progress is being made on many fronts thanks to much more precise UHECR data and the advent of multimessenger astrophysics. In this paper, we 
show how the observed spectrum and composition of UHECRs, along with 
bounds on neutrinos above 10 PeV, can be used to constrain the astrophysical properties of the environments surrounding the accelerators of UHECRs. These constraints narrow the options for candidate UHECR sources. 

As an initial demonstration of the power of this approach, we adopt an idealized description of the host environment as a sphere of size $L$ containing a uniform random magnetic field, gas, and a grey-body photon field of specified temperature. UHECR and neutrino data then point to favored ranges of temperature, and yield relations between magnetic field properties, source size, the grey-body factor and the gas column depth. 
Still more powerful constraints on the source environments will be possible when the spectrum of astrophysical neutrinos is better known and the composition of UHECRs is more accurately determined.

\section{Modeling framework}
\label{sec:framework}
This analysis is built on the Unger-Farrar-Anchordoqui framework ~\citep{UFA15} (UFA15 below), which was further explored in~\citet{MUF19} 
and significantly elaborated in~\citet{Muzio+21} (MUF19 and MFU22, respectively, to which the reader is referred for details). The basic insight of UFA15 is that the key features of the UHECR spectrum and composition --- in particular the positions of the spectral cutoff relative to the ankle and the light composition below the ankle but above the heavy, highest energy Galactic cosmic rays --- follow naturally if, after acceleration, UHECRs interact with photons or gas surrounding the accelerator, before escaping and making their journey to Earth. The critical feature of the data which demands the ``processing" of primary accelerated CRs (eschewing an ad hoc, fine-tuned separate source of protons) is the energy scale of the protonic component, which is observed to be equal to the energy per nucleon of the other components. This follows if the protons are fragments of primary CR nuclei, while if the protons were directly accelerated in the accelerator they would have the same rigidity as the other components, for a factor-of-two higher energy. Other more subtle features of the spectrum and composition give further support for the basic UFA15 picture. For specific source models which seek to explain the UHECR data see, e.g.,~\citet{Giacinti:2015pya,Globus:2015xga,Fang:2017zjf,Heinze:2019jou,Yoshida+20,Condorelli:2022vfa}.

\par
MFU22 gives an excellent description of the UHECR spectrum and composition with 8 parameters characterizing the average UHECR accelerator and its environment, and 4 nuisance parameters characterizing the highest energy Galactic cosmic rays. The accelerator is characterized by its maximum rigidity, spectral index, composition and total power in CRs per unit volume. The predictions and conclusions are quite insensitive to whether the composition emerging from the accelerator is mixed or a single $A$~(UFA15), so here we follow the fiducial model of UFA15 and treat the accelerated composition as a single $A$ to avoid introducing inessential free parameters. It was also shown~[UFA15; \citet{Fiorillo+21}] that an adequate description of UHECRs can be obtained for either a broken power-law or grey-body photon field (i.e., spectral density $n_\gamma = n_0 I_\mathrm{BB}(T)$, where $I_\mathrm{BB}(T)$ is the black-body spectral density, so $n_0 = 1$ for a black-body), with the grey-body description giving a more conservative estimate of the neutrino flux at extremely high energies~(MUF19). Here we adopt the grey-body description which avoids potentially overestimating the neutrino flux at extremely high energies due to the extended power-law tail~(MUF19) and moreover requires only two rather than four free parameters. Following UFA15, we adopt a star formation rate source evolution [SFR,~\citep{Robertson+15}], which gives among the best-fits to the UHECR spectrum~(MUF19). We show in the Appendix that our results are not strongly sensitive to the assumed source evolution.

\par
Cosmic rays interact with photons and gas until they escape the source environment. UFA15 exploited the fact that from a phenomenological perspective, what matters most in sculpting the spectrum and determining the observed composition are (1) the \emph{ratio} of escape and interaction times, (2) the peak photon energy in the source environment, and (3) how the escape time depends on rigidity. In UFA15 and MUF19, where gas in the environment was neglected, the parameters describing the environment are the temperature $T$, the ratio $r_{\rm esc} \equiv \tau^\textrm{ref}_{\rm esc}/\tau^\textrm{ref}_{\rm int} = \langle N^\mathrm{ref}_{\rm int} \rangle $ (the average number of interactions before escape for the reference nucleus), and a power-law index $\delta$ governing the rigidity dependence of $\tau_{\rm esc}$. 
Since the CR--photon cross sections and their dependence on energy and $A$ are known from laboratory experiments, interactions in the environment are fully determined once these  
parameters are specified for some reference nucleus and energy. Following UFA15, we take this reference to be $^{56}$Fe at 10~EeV. (It is immaterial whether such a nucleus is present or not in actual UHECR accelerators.)
Including interactions with gas as well as photons surrounding the accelerator~(MFU22) introduces the additional parameter $r_\mathrm{ g \gamma} \equiv \tau^\textrm{ref}_\mathrm{g}/\tau^\textrm{ref}_\gamma $.

\begin{figure*}[htpb!]
	\centering
	\begin{minipage}{0.49\linewidth}
	    \subfloat[\label{fig:fiducialfit_CR}]{\includegraphics[width=\textwidth]{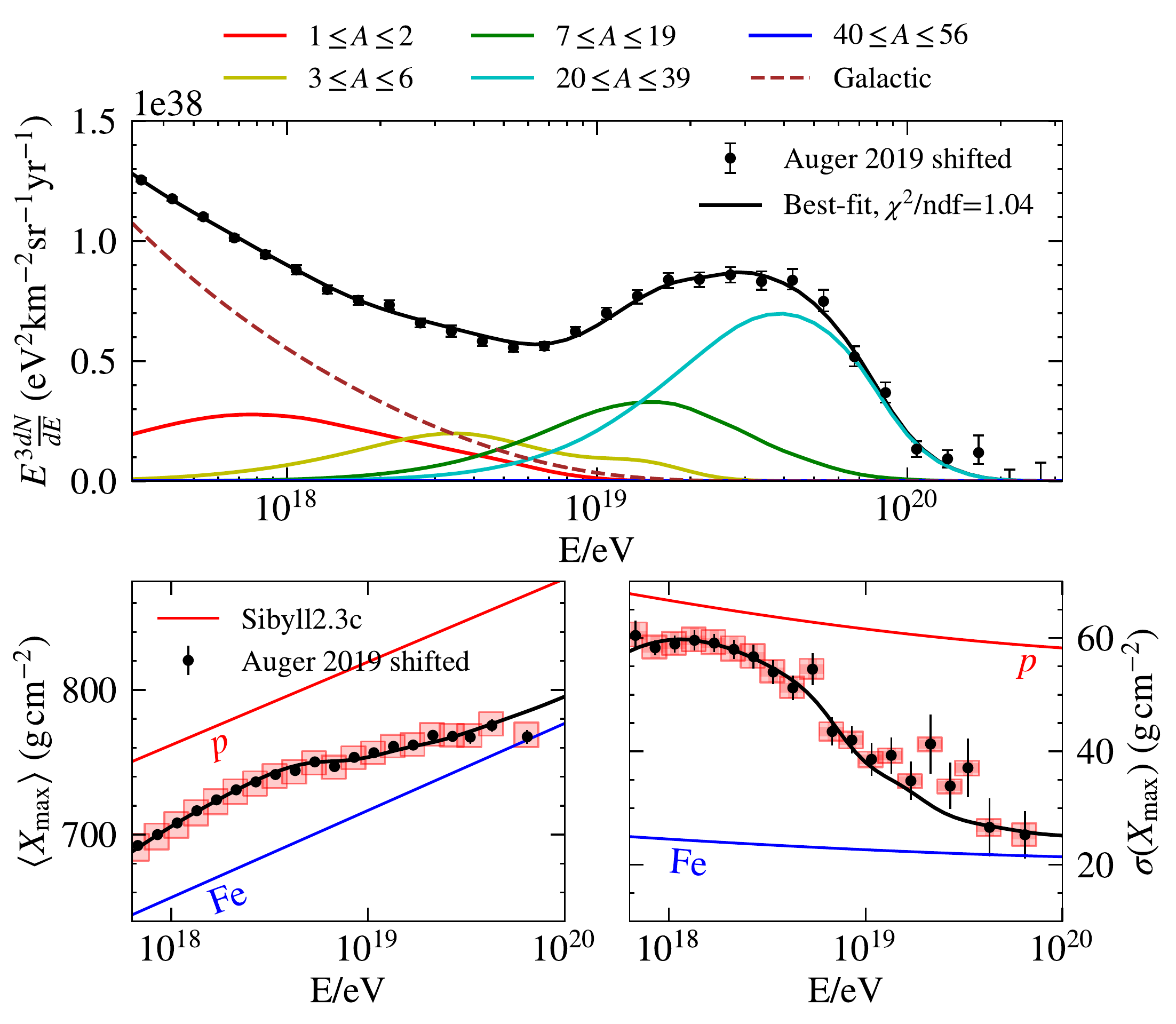}}
    \end{minipage}
    \begin{minipage}{0.49\linewidth}
        \subfloat[\label{fig:fiducialfit_neutrinos}]{\includegraphics[width=\textwidth]{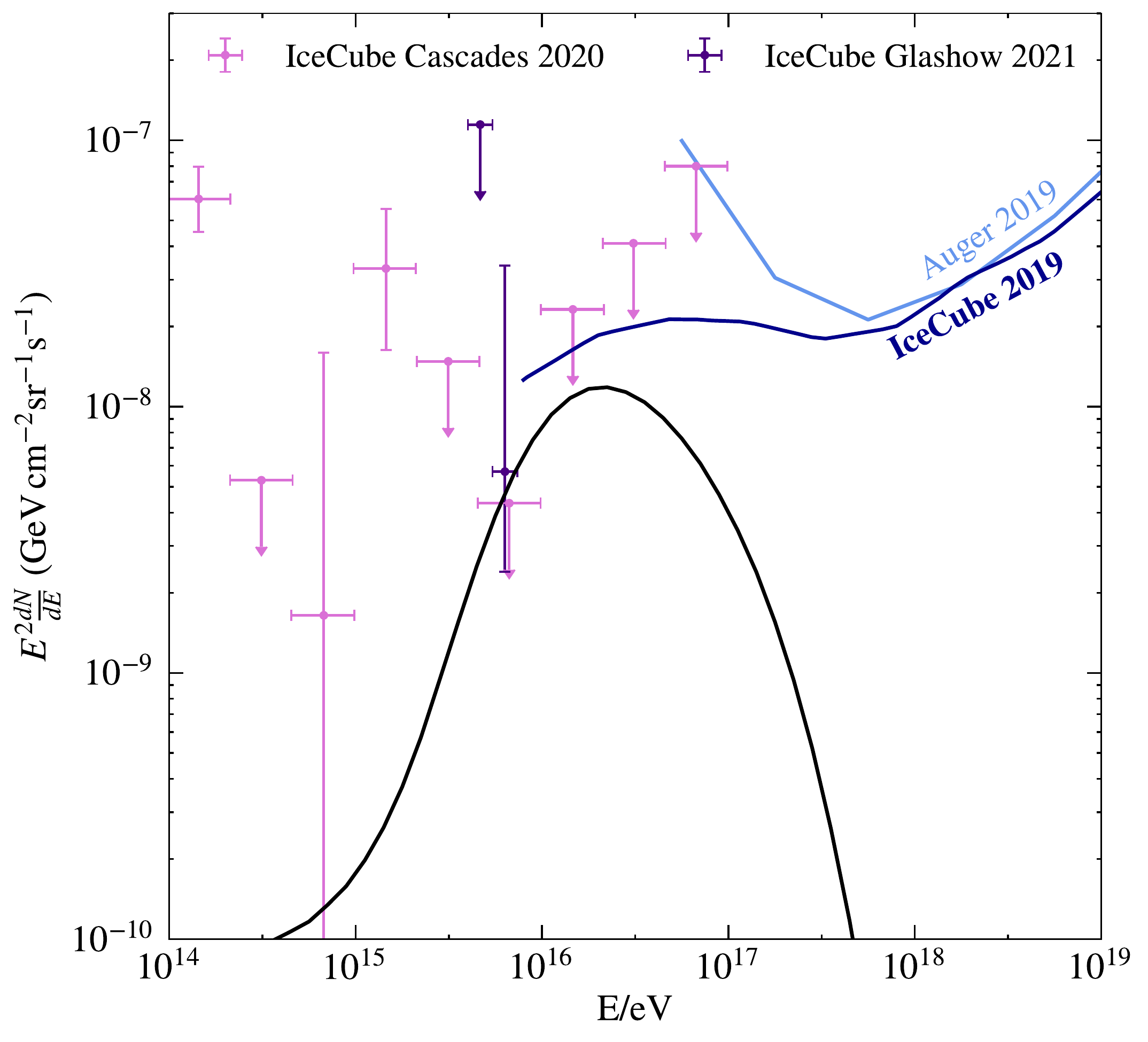}}
    \end{minipage}
	\caption{Example fit to UHECR spectrum and composition data (left; interpreted via \textsc{Sibyll2.3c}) produced by the model used in this analysis. The corresponding neutrino flux prediction (right) along with data and constraints from IceCube and Auger. Further examples of fits, including fits to the astrophysical neutrino spectrum addressed in the Appendix, can be found in~\citet{Muzio+21} (see e.g. Fig.~1 therein).}
	\label{fig:fiducialfit}
\end{figure*}

\par
An important improvement in the modeling introduced in MFU22, which we heavily exploit here, is the introduction of a more detailed description of the CR diffusion and escape, as we now discuss. 
The rate at which CRs escape, $\tau_\mathrm{esc}^{-1}$, is not in fact just a simple power-law in rigidity as in the treatment of UFA15 and MUF19. Escape depends on rigidity-dependent diffusion through a turbulent magnetic field in a source environment of characteristic size $L$. 
When the CR's Larmor radius $r_\mathrm{L}$ is much larger than the coherence length $\lambda_\mathrm{c}$, the angle of propagation changes only slightly as it crosses one coherence length: $\mathcal{O}(\lambda_\mathrm{c}/r_\mathrm{L})$. In this case, the deviation in the direction of propagation relative to the initial direction gradually increases in a diffusive manner; the CR is said to diffuse quasi-ballistically and the diffusion coefficient in distance grows as rigidity-squared, $R^2$. Instead, when $r_\mathrm{L} \ll \lambda_\mathrm{c}$, the CR direction changes completely on a scale $ \lambda_\mathrm{c}$ leading to conventional diffusion;
in this regime the spatial diffusion coefficient is much smaller than in the quasi-ballistic regime and has a different functional dependence on rigidity. 

Taking the turbulence to be isotropic Kolmogorov and defining a dimensionless diffusion coefficient $d(R)$ such that $D(R) \equiv c \lambda_\mathrm{c} d(R)/6\pi$, tracking simulations are well-fit by
\begin{equation}
d(R) = \left(\frac{R}{R_\mathrm{diff}}\right)^{1/3} + \frac{1}{2} \left(\frac{R}{R_\mathrm{diff}}\right) + \frac{2}{3} \left(\frac{R}{R_\mathrm{diff}}\right)^2~,\label{eq:dr}
\end{equation} 
where $R_\mathrm{diff}$ is the rigidity at which the Larmor orbit equals the coherence length of the turbulent magnetic field: $2\pi r_\mathrm{L}(R_\mathrm{diff}) \equiv \lambda_\mathrm{c}$. (The coefficients of the various terms in (\ref{eq:dr}) come from our fit to the tracking results reported in~\citet{Globus+07} and are only accidentally adequately approximated as simple fractions; see MFU22 for details.) The change in slope of the power-law behavior of CR propagation, in the rigidity range such that $r_\mathrm{L} \approx \lambda_\mathrm{c}$, leaves an imprint on the UHECR spectrum and composition which is sensitive to the magnetic field properties.  This is especially constraining if $R_\mathrm{diff}$ is in the rigidity range of the UHECR data, as proves to be the case. We exploit this here to constrain $B$ and $\lambda_\mathrm{c}$. It should be noted that even if $R_\mathrm{diff}$ were outside the UHE range and its value could not be determined from fitting UHECR data, the slope of the power-law behavior of CR propagation would still indicate whether $R_\mathrm{diff}$ is above or below the rigidity range of the UHECRs and place a bound on $R_\mathrm{diff}$.

In MFU22 the escape time is modeled as
\begin{align} \label{eq:escTime_def}
  \tau_\mathrm{esc}(R) = \frac{L^2}{6D(R)} + \frac{L}{c} ~.
\end{align}
The escape time can be written in terms of the escape time of the reference nucleus $\tau_\mathrm{esc}^\mathrm{ref}$ as
\begin{align}
  \tau_\mathrm{esc}(R) = \tau_\mathrm{esc}^\mathrm{ref} \left( \frac{\pi \, r_\mathrm{size}}{d(R)} +1\right)  \left(\frac{\pi \, r_\mathrm{size}}{d(R_\mathrm{ref})} +1\right)^{-1},
\end{align}
where $R_\mathrm{ref} \equiv 10/26$~EV $\simeq 0.38$~EV is the rigidity of the reference nucleus and the model parameter $r_\mathrm{size} \equiv L/\lambda_\mathrm{c}$ is the size of the environment in units of the coherence length of its random magnetic field.

\section{Analysis and Results}

\par
Our results are based on the MFU22 analysis framework that uses
the algorithms described in~UFA15 for a fast evaluation of the composition and spectra at Earth given the parameters of the sources and their environment; details are given in~MFU22. The strongest constraints come from the Auger UHECR spectrum and composition-sensitive observables $\langle X_\mathrm{max} \rangle$ and $\sigma\left(X_\mathrm{max}\right)$~\citep{Aab+20a,Aab+20b,Verzi20,Abreu+13,Aab+14a,Aab+14b,Yushkov20}.  Our analysis could be applied to Telescope Array (TA) spectrum and composition data~\citep{Bergman:2021zmo,Zhezher:2021qke}, however we use the Auger data since Auger's larger exposure allows for higher statistics measurements of $X_\mathrm{max}$, and moreover the observations made by both observatories agree within systematic uncertainties over most of the energy range~\citep{TelescopeArray:2021zox}.  

\par
Interpretation of the $X_\mathrm{max}$ observables in terms of composition requires a hadronic interaction model (HIM), for which we use both \textsc{EPOS-LHC}~\citep{Pierog+13} and \textsc{Sibyll2.3c}~\citep{Fedynitch+18}, to assess the sensitivity of our results to the HIM. As we shall see, the conclusions are insensitive to the choice of HIM.

\begin{figure}
	\centering
    \includegraphics[width=\linewidth]{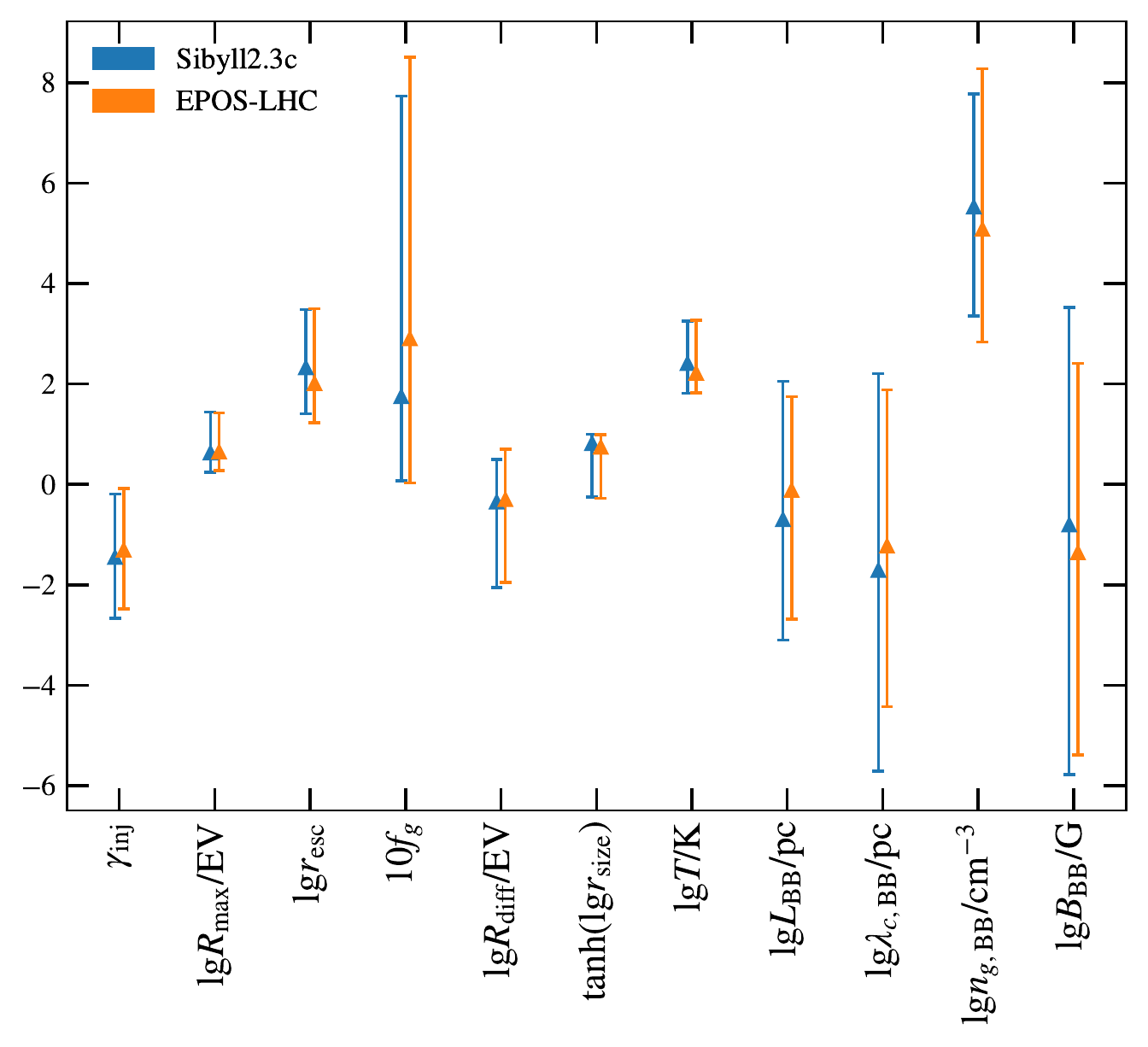}
    \vspace{-0.3in}
	\caption{Ranges for some key model and astrophysical parameters derived from this analysis, reporting results for $L$, $\lambda_\mathrm{c}$, $n_\mathrm{g}$, and $B$ for a black-body ($n_0=1$) photon field; the conversion for other $n_0$ values is given in the text.
	Central values indicate the median of the posterior distributions while error bars indicate the $16$th and $84$th percentiles (i.e., these are not best-fit values and error bars on the parameters of a particular model). The results for the \textsc{Sibyll2.3c} and \textsc{EPOS-LHC} hadronic interaction models are shown in blue and orange respectively. Due to correlations between parameters, certain combinations are better constrained than the overall allowed range of individual parameters might suggest, as can be seen from the corner plots showing the joint probability distribution of pairs of parameters in Appendix~\ref{app:cornerplots}.}
	\label{fig:constraints_comparison_CR}
\end{figure}

\par
As discussed in~MFU22, constraints from the extragalactic gamma-ray background reported by \textit{Fermi}-Large Area Telescope (LAT)~\citep{Ackermann+14} are presently weaker than, and fully captured by, the constraints imposed by the IceCube bounds on neutrinos above $10^{15.9}$~eV. Gamma-rays at $\gtrsim$~TeV energies do not currently constrain UHECR sources as the predicted flux is steeply falling at these energies (see Fig. 1 of~MFU22). Since gamma-rays generally are not currently constraining we omit them for simplicity. 

\par
We perform a Markov chain Monte Carlo (MCMC) exploration of the 12-dimensional parameter space with each HIM. This MCMC analysis was carried out using \verb|emcee|~\citep{ForemanMackey+12}, fitting UHECR data and rejecting models which predict $N_\nu > 4.74$ above $10^{15.9}$~eV at the $99\%$~CL~\citep{Feldman+97}, as this violates bounds on extremely high energy (EHE) neutrinos from IceCube~\citep{Aartsen+18,IceCubeGlashow2021}. 

To understand the impact of the neutrinos on our conclusions, we report in the Appendix the results of fitting only the UHECR data without neutrino constraints, or fitting the high energy neutrino data points as well as the UHECR data.  The best fit turns out to be the same when fitting just the UHECR data or imposing the neutrino upper limits, but the shape of the posterior distribution is somewhat different. Actually fitting to both the UHECR and the neutrino data is not significantly different from the fiducial model using just the neutrino upper limits.  We choose to use just the neutrino upper limits for our fiducial model because the astrophysical neutrino spectrum is still fairly uncertain and different analyses give rather different spectra, so it would not be clear which to adopt. 

\begin{figure*}[htpb!]
	\centering
	\begin{minipage}{0.49\linewidth}
	   \subfloat[\label{fig:BL_constraints_CR_sibyll}]{\includegraphics[width=\textwidth]{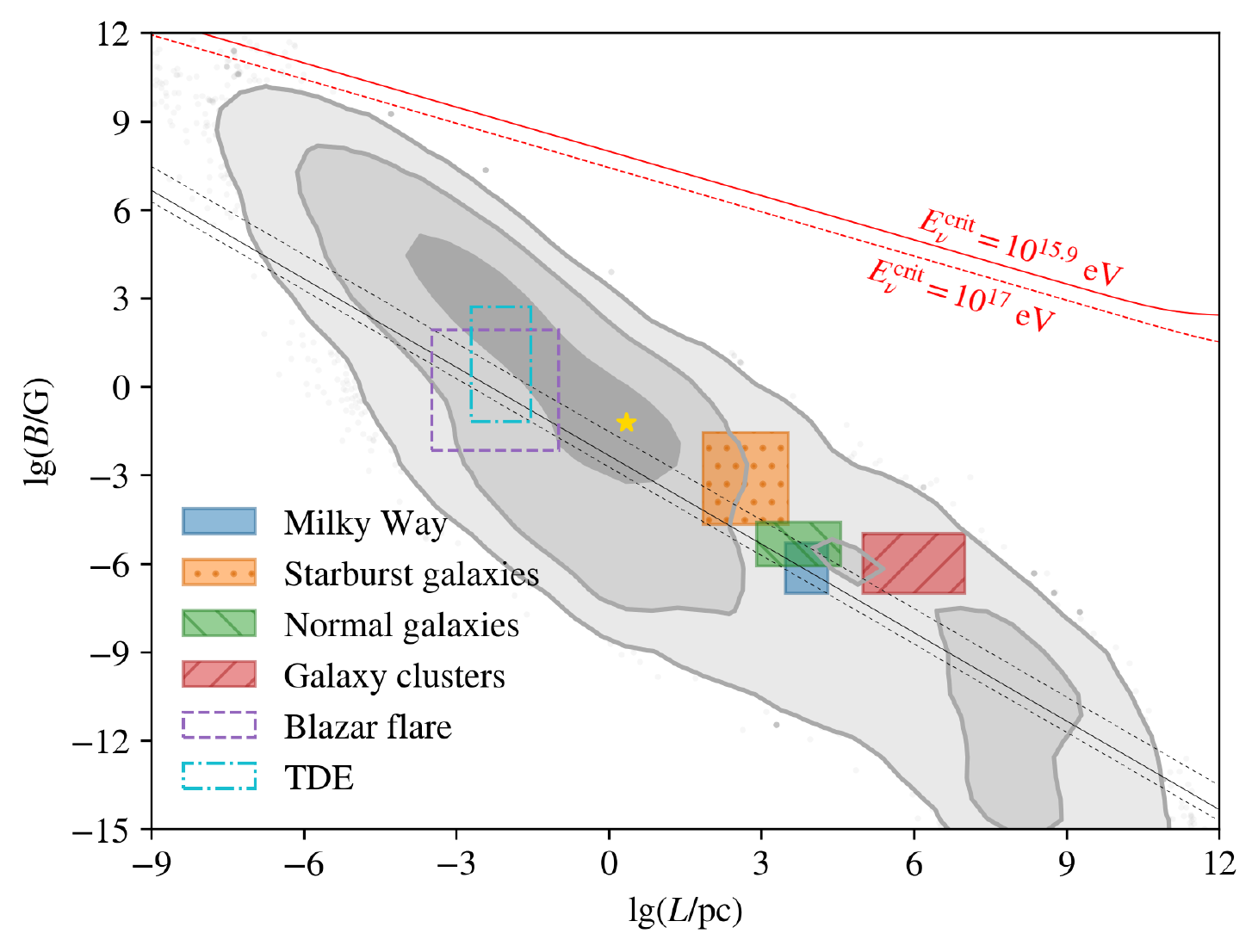}}
    \end{minipage}
    \begin{minipage}{0.49\linewidth}
      \subfloat[\label{fig:BL_LnGas_constraints_CR_sibyll}]{\includegraphics[width=\textwidth]{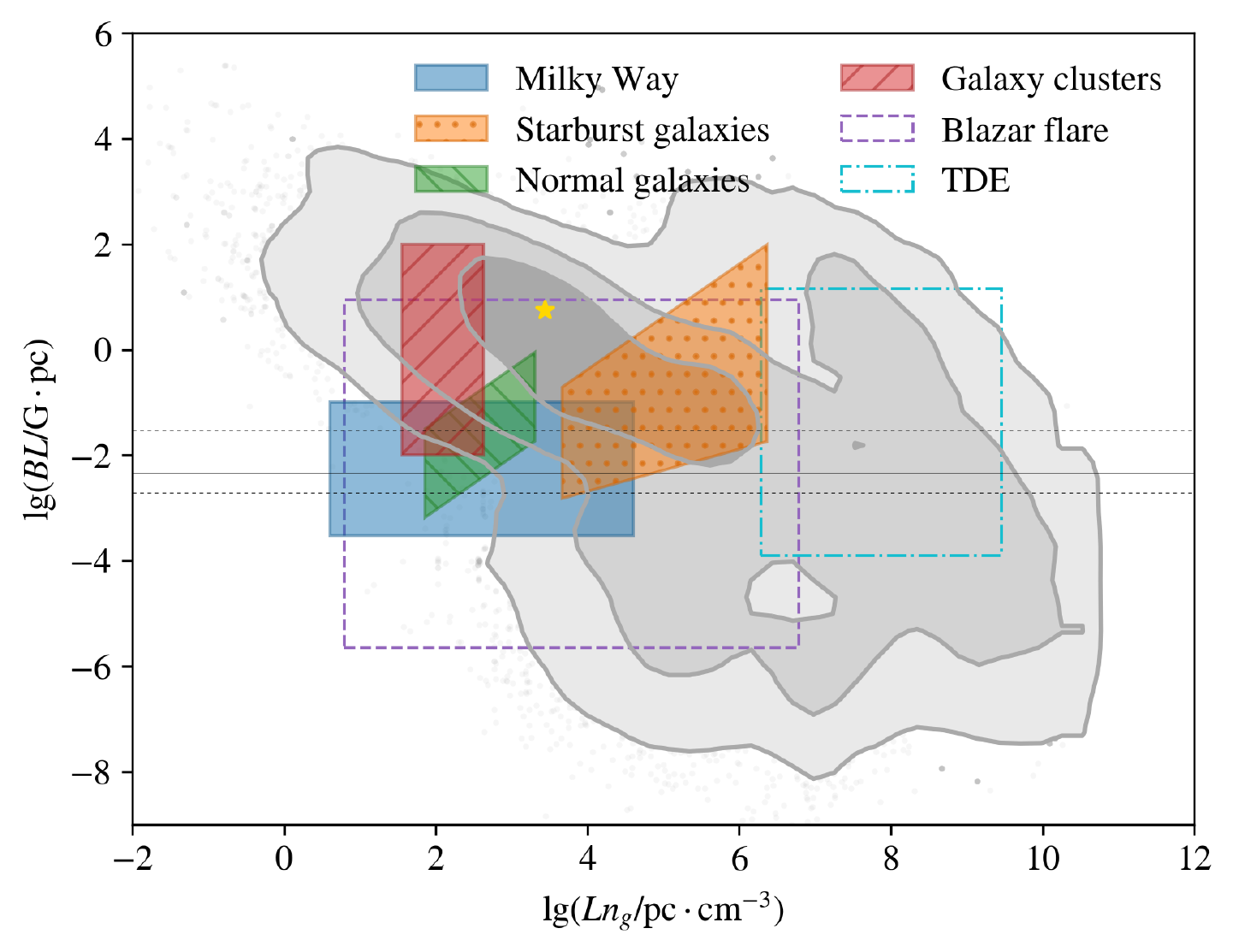}}
    \end{minipage}
	\caption{{\bf Left:} The grey regions show the posterior probability distribution as a function of effective size $L$ and magnetic field strength $B$ of the source environment, 
	using \textsc{Sibyll2.3c} and taking a black-body spectrum, $n_0 = 1$. The shaded regions give the $1\sigma$, $2\sigma$, and $3\sigma$ uncertainty bands (darkest to lightest grey, respectively) of the joint posterior distribution. The peak of the distribution 
	is indicated by the gold star. For $n_0 \neq 1$, the posterior distribution shifts according to $B= n_0B_\mathrm{BB}$ and $L=L_\mathrm{BB}/n_0$. 
	The solid black diagonal line shows, for reference, $B_{\rm acc}$ vs $ L_{\rm acc}$ \textit{in the accelerator} such that the Hillas criterion is satisfied for particles emerging from the accelerator at the median rigidity of the posterior distribution; the dashed lines show the same for the $16$th/$84$th percentiles. The red lines demarcate regions where synchrotron losses become significant; see Appendix~\ref{app:syncCooling} for details.
	The approximate range of size and magnetic field strength of various potential source types are indicated in shaded boxes as a guide. Recent multimessenger candidate sources of neutrinos are shown by dashed boxes, as more examples are needed for the correlation to be substantiated. {\bf Right:} The joint posterior probability distribution for $BL$ and $L n_\mathrm{g}$, the surface number density of gas in the source environment. These products are independent of the value of $n_0$ so that this joint posterior distribution is unaffected by the value of $n_0$.}
	
	\label{fig:srcs_constraints_CR}
\end{figure*}

\par
An example fit from our analysis, fitting the UHECR data subject to EHE neutrino constraints, is shown in Fig.~\ref{fig:fiducialfit}.   Fig.~\ref{fig:constraints_comparison_CR} displays the posterior parameter ranges for individual parameters. Fit parameters that depend on the grey-body factor $n_0$ are reported for the black-body case $n_0=1$; the conversion for other $n_0$ values is $L = L_\mathrm{BB}/n_0$, $B = n_0 \, B_\mathrm{BB} $, $\lambda_\mathrm{c} = \lambda_\mathrm{c,BB} / n_0$, and $n_\mathrm{g} = n_0~ n_{g,\mathrm{BB}}$.  One sees from Fig.~\ref{fig:constraints_comparison_CR} that most parameters are insensitive to the underlying HIM assumed. Parameter values and corner plots for all of the data variations explored and for both HIMs are given in the Appendix.   

\section{Astrophysical constraints}

\par
Constraints on the model parameters can be translated into constraints on astrophysical parameters. 
A powerful result of this analysis is the clear preference for a low temperature source environment (see Fig.~\ref{fig:constraints_comparison_CR}), which disfavors a number of otherwise attractive source candidates. 
UHECR data alone does not discriminate well between gas- or photon-dominated interactions, although it shows a slight preference towards the latter~(MUF19). However the fraction of CR interactions with gas is highly constrained by limits on the number of EHE neutrinos, demanding a significant fraction of the source interactions be with photons -- with the consequence that the photon temperature is well constrained to be relatively cool, $\mathcal{O}(100-1000)$~K.  As the temperature increases beyond $\approx 1000$~K, two effects contribute to a dramatically increasing rate of pion and hence neutrino production: the center-of-mass energy of the CR-photon interaction increases, and the number density of photons increases like $T^3$.  At significantly higher temperatures, nuclei are entirely destroyed and a fit to the UHECR composition data is impossible, unless the photon field around the source is unphysically thin.

\par
While some model parameters are directly astrophysical parameters, such as the photon field temperature and maximum rigidity of the accelerator, other model parameters provide constraints on relationships between parameters of the source environment.
Some key relationships are:

$\bullet$
$r_\mathrm{esc}$ is the ratio of the escape and interaction times of the reference nucleus with gas and photons, where $\tau_\mathrm{int}^{-1} = \tau_\mathrm{g}^{-1} + \tau_{\gamma}^{-1}$, and $r_\mathrm{ g \gamma}$ fixes the ratio $\left. \tau_\mathrm{g}^\mathrm{ref}\middle/\tau_\gamma^\mathrm{ref} \right.$. Combining these definitions and using (\ref{eq:escTime_def}) gives
\begin{equation}
\label{eq:n0L}
  n_0 \, L = \frac{c \,\tau_{\mathrm{BB},\gamma}^\mathrm{ref} \, r_\mathrm{esc} \, r_\mathrm{g\gamma}}{\left(\pi \, r_\mathrm{size}\middle/d(R^\mathrm{ref})+1\right)\left(1+r_\mathrm{g\gamma}\right)},
\end{equation}
where $\tau_{\mathrm{BB},\gamma}^\mathrm{ref}$ is the total photohadronic interaction time for the reference nucleus with a black-body photon spectrum of temperature $T$, and $n_0$ is the dimensionless grey-body scaling factor.

$\bullet$ From the definition of $R_\mathrm{diff}$ 

\begin{align} \label{eq:Blambdac}
  B\lambda_\mathrm{c} \simeq 2.2\pi\left( \frac{R_\mathrm{diff}}{\text{EV}} \right) \text{$\mu$G$\cdot$kpc}.
\end{align}

\noindent
Fitting the UHECR spectrum and composition constrains $R_\mathrm{diff}$, and therefore places a constraint on the turbulent magnetic field in the source environment as discussed below Eq.~\eqref{eq:dr}.

%\\ 
$\bullet$
The fit also fixes $r_\mathrm{g\gamma}$, determining the relationship between the gas density and grey-body scaling factor in the source. From the definition $\tau_\mathrm{g}^{-1}(E,A) \equiv n_\mathrm{g} \sigma_\mathrm{g}(E, A) c$: 
\begin{align}
\label{eq:ph2gasRatio}
  \frac{n_0}{n_\mathrm{g}} = r_\mathrm{g\gamma}\, c \, \tau_{\mathrm{BB},\gamma}^\mathrm{ref}(T)\, \sigma_\mathrm{g}^\mathrm{ref}~.
\end{align}

Using (\ref{eq:ph2gasRatio}), the constraints (\ref{eq:Blambdac},~\ref{eq:n0L}) can be combined in multiple ways, depending on the information available for a particular candidate source. Eq. (\ref{eq:n0L}) constrains the product of the effective size of the source environment and the intensity of the photon field, but since $r_\mathrm{size} = L/\lambda_\mathrm{c}$ is a parameter of the fit, the source size $L$ can be eliminated to write relations in terms of intrinsic features, $B,~\lambda_\mathrm{c}$ and $n_0$. 

\par
Joint posterior distributions between parameters can thus be obtained from our analysis using the results of Sec.~\ref{sec:framework} and those above. For example, to obtain the joint posterior distribution between $B$ and $L$ 
we note that $B \times L$ is fixed by $r_\mathrm{size}$ and (\ref{eq:Blambdac}), while the value of $L$ is fixed by (\ref{eq:n0L}) for a given grey-body factor $n_0$. Marginalizing over all other parameters from our MCMC analysis, we obtain the joint posterior distribution between $B$ and $L$ for a given value of $n_0$. 
This is shown in Fig.~\ref{fig:srcs_constraints_CR} for $n_0=1$ using \textsc{Sibyll2.3c}.  For $n_0 \neq 1$, the posterior distribution shifts according to $B= n_0B_\mathrm{BB}$ and $L=L_\mathrm{BB}/n_0$. Corner plots showing the joint posterior distribution between other astrophysical source properties are given in the Appendix~\ref{app:cornerplots}.

\par
In the derivation of (\ref{eq:n0L}) and (\ref{eq:Blambdac}) and in defining $r_\mathrm{esc}$, we assumed the region of magnetic confinement was the same as the region containing the gas and/or photons where most interactions occur. This is obviously an idealization and could be elaborated further. But our analysis applies equally well if the region of magnetic confinement around the source extends beyond $L$, the interaction region, because an increased UHECR pathlength due to magnetic deflections outside the region $L$ does not impact the multimessenger data (unless the additional propagation length materially extends the UHECR's propagation in the extragalactic photon field; in that case the effect factorizes and can be treated separately~\citep{Harari+16}). If the magnetic confinement region extends beyond the interaction region, $\lambda_\mathrm{c} $ can exceed $ L$ as may be relevant for some cases.

\section{Interpretation}
Figure~\ref{fig:srcs_constraints_CR} shows (colored boxes) the approximate ranges of $B$ and $L$ characteristic of several potential UHECR accelerator hosts and other benchmark systems, from the literature, superimposed on posterior distributions from our analysis. For the Milky Way the domain shown is based on parameters given in~\citet{Jansson+12} and~\citet{Kennicutt+12}; for starburst galaxies (SBGs) and normal star-forming galaxies we followed~\citet{Thompson+06}; for galaxy clusters the region is based on parameters inferred in~\citet{Ptitsyna+08} and observations from~\citet{Croston+08}. In addition to the classic candidates for UHECR sources, the dashed boxes show two transient possibilities, TXS 0506+056~\citep{TXSObservation} and TDE AT2019dsg~\citep{Stein+20} based respectively on the multimessenger studies in~\citet{Cerruti+18,Keivani+18,Liu+18,Murase:2018iyl,Gao+18,MAGIC18,Xue+19,Zhang+19} and~\citet{Stein+20,Liu+20,Murase:2020lnu,Winter+20,Cendes+21}. Their large ranges reflect both the uncertainties in the interpretation of the observations and the potentially large inherent range of conditions. The box for a given system is inclusive in the sense that regions exhibiting the given $B$ can be found, with $L$ in the range shown, but not every combination of $B$ and $L$ within the colored box may be realized in the system. Refining these domains to distinguish the properties of particular sub-regions of candidate sources and their surroundings, e.g., the base of an AGN jet versus the external shock at the radio lobes, is needed in order to fully exploit our constraints.  

\par
The black lines in the left panel of Fig.~\ref{fig:srcs_constraints_CR} show the Hillas criterion for the accelerator: the locus of $B_\mathrm{acc}L_\mathrm{acc}$ such that the Larmor radius of the maximum energy CRs equals the size $L_\mathrm{acc}$ of the accelerator. Since our fit to the UHECR data determines the rigidity distribution of the UHECRs emerging from the accelerator, this is a more exact representation of the Hillas criterion than the usual band taking CRs to have charge somewhere between $Z=1$ and $Z=26$. There is no a priori relation between $B L$ in the environment and $B_\mathrm{acc}L_\mathrm{acc}$, but their ratio gives an indication of the source environment's properties compared to those of the accelerator. For example, if the magnetic field in the accelerating region is of comparable strength to that in the interaction region, then this ratio is the size of the source environment relative to the size of the accelerator. Our results favor this ratio to be in the range of $\sim 1$ to $\sim 10^5$, with median $\sim 10^2$; this provides an additional potential probe of UHECR sources.

\par 
To use the constraints embodied in the left panel of Fig.~\ref{fig:srcs_constraints_CR} requires knowing the grey-body factor $n_0$ of the photon field. 
For systems which are approximately black-body, 
the posterior distribution in Fig.~\ref{fig:srcs_constraints_CR} can be used directly, but otherwise $n_0$ must be determined, which can be non-trivial. For example, based on results of~\citet{Liu+18} for the broad-line region of TXS 0506+056, 
$n_0 \approx 10^{-4.7}$. For this value, the posterior distribution would be obtained from the one for $n_0 = 1$ by sliding it downward and to the right parallel to the ``Hillas rails" by $10^{-4.7}$ and $10^{+4.7}$, respectively. If this $n_0$ estimate and the box in the $B-L$ plane attributed to TXS 0506+056 are valid, TXS 0506+056 would be strongly disfavored as a source of UHECRs. 

\par
The right panel of Fig.~\ref{fig:srcs_constraints_CR} provides a complementary set of constraints on source properties, independently of the value of $n_0$. Here, we frame the constraints in terms of $BL$ and $L n_\mathrm{g} \simeq \Sigma_\mathrm{g}/m_p$, the surface number density of gas, using Eqs. (\ref{eq:Blambdac})-(\ref{eq:ph2gasRatio}). 
The constraints shown in this plot are independent of and complementary to the constraints in the left panel;  they are especially valuable for cases where $n_0$ cannot be readily determined. The colored boxes for different candidates are large here, because within a given system different potential accelerator loci are surrounded by quite different environments. This just means that more refined decomposition into conditions in specific loci of the systems is needed to fully exploit our constraints, by replacing the large boxes with much more circumscribed domains, some of which will be excluded. 

\par
Another general constraint on the interaction region itself is the fit parameter $r_\mathrm{size} \equiv L/\lambda_\mathrm{c}$. Although the uncertainties on this quantity are large within our current analysis framework (see Fig.~\ref{fig:constraints_comparison_CR}), future more specialized modeling could reduce the uncertainties.  Tables of all fit results are given in the Appendix~\ref{app:preferredparams}.

\section{Some applications}
The simplified treatment given here assuming a homogeneous source environment, is a good approximation for some source candidates but not for all. If the simple treatment is applicable, the region of the source environment responsible for the bulk of CR interactions should have properties consistent with the high-posterior region obtained in this analysis. It is insufficient to have compatibility with some properties, e.g. magnetic field strength and source size, if another property, e.g. temperature, is far from the peak of the posterior distribution. The requirement that a system lie within the favored region for all constrained parameters imposes a strong condition on candidate sources. Corner plots in  Appendix~\ref{app:cornerplots} detail the interplay between major source properties. Only a subset of the constraints -- on temperature, size and magnetic field -- are employed in this first analysis.

One proposed UHECR source type which our analysis appears to decisively rule out, is acceleration in the large scale shocks of massive galaxy clusters~\citep{blandford+ClusterAccel18}. The gas temperature in these systems, also called X-ray clusters, is $\mathcal{O}(10^7-10^8)$~K; the observed X-rays {\it are} the black-body photons. Hence, the temperature is much higher than is compatible with our constraints. Moreover since the photon field in the cluster is a black-body, $n_0=1$ and Fig.~\ref{fig:srcs_constraints_CR} shows that the \{B, L\} values are far from the favored region. Conceivably a domain in the outer, cooler region of clusters can have $T$ small enough, while satisfying the \{B, L\} requirements for the relevant $n_0$ value. However massive galaxy clusters have an additional challenge as far as being the sources of UHECRs:  the UFA mechanism's successful explanation of the sub-ankle composition and spectrum relies on the acceleration stage being completed \emph{before} the UHECRs are subjected to interactions in the surroundings~(UFA15).  The possibility that acceleration occurs in shocks at the surface of X-ray clusters, with the UHECRs escaping the cluster environment without being processed and then being processed during travel through cosmic filaments en route to the Galaxy or in the Galactic halo, is not viable because filaments and the halo produce by-far-insufficient processing. 

\par
Starburst Galaxies (SBG, also known as Luminous Infrared Galaxies) were identified as a possible UHECR source in~\citet{bfzLIRG10}.   The evidence for a directional correlation between SBGs and UHECRs was strengthened in~\citet{augerSBG18}, albeit without taking into account coherent magnetic deflections in the Galaxy. The relevant parameters of typical and extreme SBGs (exemplified by M82 and Arp 220) are determined in Appendix~\ref{app:SBGs} by fitting their SEDs. Both have a similar temperature, at the low end of the fit range. Arp 220 has $n_0\approx 1$ so Fig.~\ref{fig:srcs_constraints_CR} applies directly, showing that extreme SBGs like Arp 220 cannot be major sources of UHECRs. However M82 has $n_0\approx 10^{-2}-10^{-3}$, sliding the posterior distribution 2-3 units to the right and down, for good agreement with the $B$ and $L$ range estimated for SBGs.

\section{Summary}
We have used a flexible phenomenological model of UHECR sources and their surroundings, developed in~UFA15 and elaborated in~MUF19 and MFU22, to constrain properties of the UHECR source environment consistent with up-to-date multimessenger data. Our treatment is agnostic to the exact acceleration mechanism and the particular astrophysical source of UHECRs, yet enables us to extract powerful information on source properties. 
UHECR and neutrino data reveal a consistent picture of the preferred astrophysical properties of UHECR sources -- whether simultaneously fitting astrophysical neutrino data or only imposing consistency with bounds on EHE neutrinos. There is little sensitivity to the hadronic interaction model. 

\par 
In general, significant UHECR interactions may occur in various regions of the source environment. It is the cumulative effect of these regions which matters, but for simplicity in this initial paper we imagine that only one homogeneous region accounts for most of the interactions.
For such systems, our results show that after UHECRs escape from their accelerator they pass through and interact with a photon field whose black-body-equivalent temperature is $\mathcal{O}($100-1000)~K. If this region is black-body, it is small -- $\lesssim 100$~pc -- and its RMS magnetic field strength is $\gtrsim 100 \, \mu$G, suggestive of compact systems like TDEs and some parts of AGNs. But another possibility is that the photon field is a low-density grey-body with $n_0 \ll 1$, of larger size and weaker magnetic field. Typical starburst galaxies are viable source candidates of the second type, but ultrahigh luminosity SBGs like Arp 220 have an approximately black-body photon field which is incompatible with the constraints; hence those cannot contribute a major component of observed UHECRs. The suggestion that UHECR acceleration occurs in the large scale shocks of galaxy clusters seems to be ruled out by our constraints.

\par
The approach taken in this paper is complementary to other, more tailored studies of specific source candidates.  Our results are in good agreement with \citet{Keivani+18} who conclude that multimessenger data make it unlikely for TXS-0506+056 to be a UHECR accelerator.  The recent study of \citet{Condorelli:2022vfa} on SBGs as potential sources of UHECRs, which appeared subsequent to the posting of our paper on the arXiv, agrees with our conclusions.  Other candidate source types, e.g., AGN, are so complex that the overall system comprises multiple regions, so comparison of our results to source-specific studies are more difficult. For such systems, our approach can be tailored to incorporate the locus of the accelerator within the system and known photon spectra in different regions and detailed system geometry. This will help differentiate which particular acceleration regions are acceptable, or perhaps exclude an entire source type. 

\par
Application of the results presented here should help to identify the most promising candidates for the accelerators of UHECRs for further work.  While focused source studies like those cited above are useful for understanding the challenges particular sources face in explaining UHECR data, our methodology allows for a broad assessment of which candidate UHECR sources are viable.

\begin{acknowledgments}

We are indebted to Michael Unger for his invaluable input; we thank Todd Thompson for helpful information about conditions in starburst galaxies and other systems, and Foteini Oikonomou for useful feedback on our analysis. The research of MSM was supported in part by the NYU James Arthur Graduate Award, the Ted Keusseff Fellowship, and the NSF MPS-Ascend Postdoctoral Award \#2138121. The research of MSM and GRF was supported in part by NSF-2013199. This work was supported in part through the NYU IT High Performance Computing resources, services, and staff expertise.

\end{acknowledgments}

\appendix

\section{Overview of analysis cases}\label{app:overview}

\par
We performed an MCMC exploration of the 12-dimensional model parameter space for two hadronic interaction models (HIMs), \textsc{Sibyll2.3c} and \textsc{EPOS-LHC}, and considering three nested cases: (1) fitting UHECR data alone, (2) fitting UHECR data alone but rejecting models which violate the IceCube neutrino bounds at the $99\%$ CL, and (3) simultaneously fitting Auger UHECR and IceCube astrophysical neutrino data. We consider the case fitting UHECR data subject to IceCube neutrino constraints to be our fiducial case, and it is the focus of the Letter. The case in which we simultaneously fit the UHECR and astrophysical neutrino data makes the additional assumption of a common origin of UHECRs and the high energy portion of the astrophysical neutrino spectrum. The case fitting UHECR data alone should only be considered as illustrative: comparing it to the other cases shows the effect of EHE neutrino constraints on the results but this case is not an acceptable model, since neutrino constraints must be respected in an analysis of UHECR sources.

\par
For the case simultaneously fitting UHECR and astrophysical neutrino data, the sum of the $\chi^2$ for the UHECR data and the $\chi^2_\nu$ for the astrophysical neutrino data is used in the likelihood function. We include a low-energy neutrino component to supplement the UHECR-produced component, parametrized as a single power law with an exponential cutoff.
We calculate a $\chi^2_{\nu,0}$ to the data points of the IceCube Glashow event observation~\citep{IceCubeGlashow2021} and to the IceCube Cascades data set between $16$~TeV and $2.6$~PeV, the sensitive range for the Cascades analysis as determined by IceCube~\citep{IceCubeCascades20}. Upper-bounds are included by adding $2n_i$ to the $\chi^2_{\nu,0}$, where $n_i$ is the expected number of events predicted by the model in energy bin $i$~\citep{Baker+83}, so the final measure of the neutrino goodness-of-fit is given by $\chi^2_\nu = \chi^2_{\nu,0} + 2\sum_i n_i$, where $i$ runs over energy bins with upper-bounds.

\par
We note that the specific value of the $\chi^2$ is not particularly meaningful for this analysis due to the dominance of systematic uncertainties over statistical in most data points. However, the difference between $\chi^2$'s is well-defined, so that fits are well-constrained, as is most important for this analysis. Overall the $\chi^2/ndf$ is in the $1-2$ range for the best-fits depending on the specifics of the model (see MFU22 for details).

\section{Preferred parameter values}\label{app:preferredparams}

\par
\begin{figure}
	\centering
    \includegraphics[width=\linewidth]{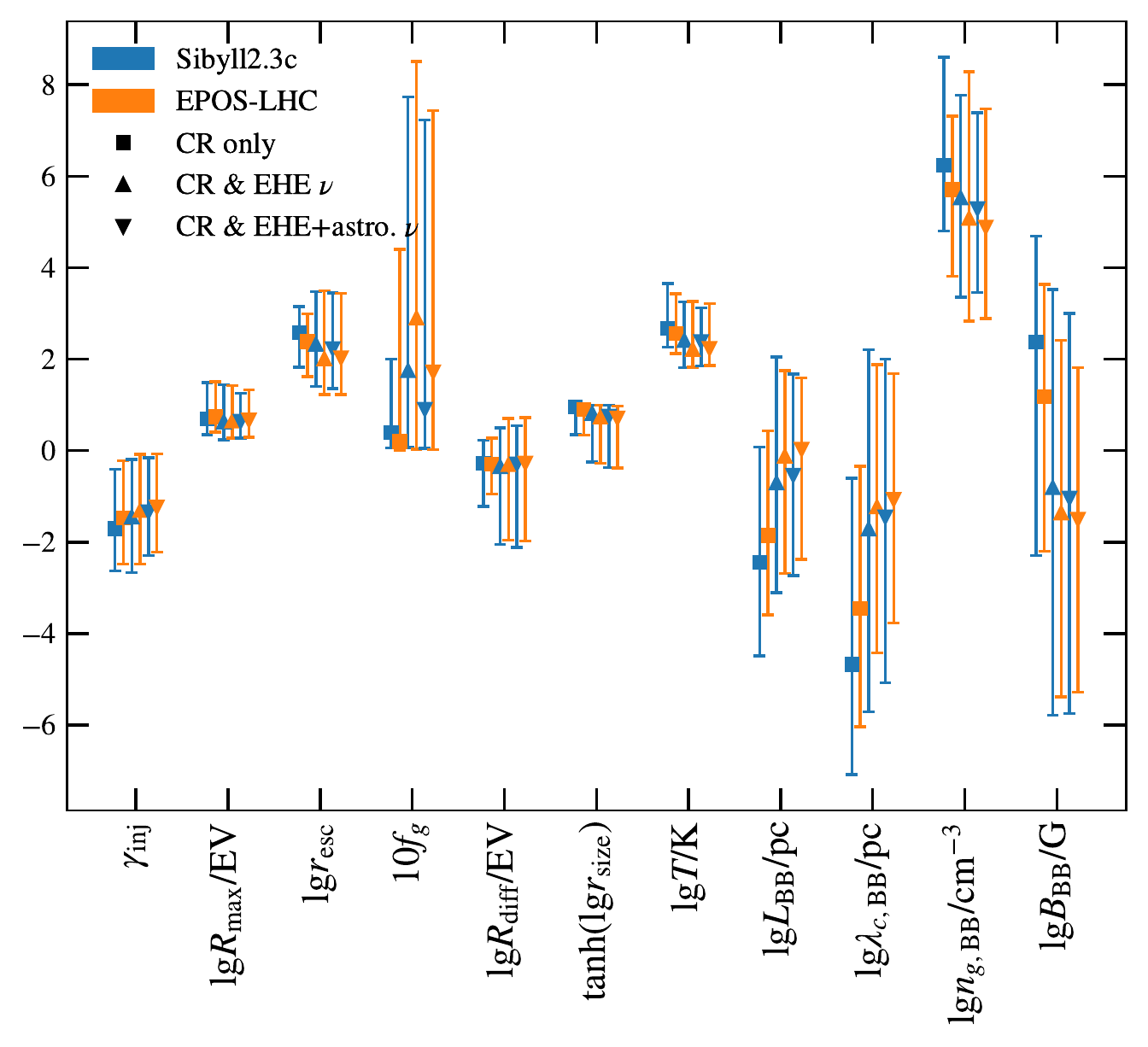}
    \vspace{-0.3in}
	\caption{Comparison of model and astrophysical parameters fitting CR data alone (squares), rejecting models violating EHE neutrino bounds (upward triangles), and simultaneously fitting astrophysical neutrino data (downward triangles) using the \textsc{Sibyll2.3c} (blue) or \textsc{EPOS-LHC} (orange) HIMs. Central values indicate the median while error bars indicate the $16$th and $84$th percentiles of the posterior distributions. The last four parameters depend on the choice of $n_0$ and are shown for $n_0=1$; the scaling for $n_0 \neq 1$ is given in the text.}
	\label{fig:constraints_comparison}
\end{figure}

In this section we report 
the results of three analysis cases, for the two HIMs. Figure~\ref{fig:constraints_comparison} shows a direct comparison of astrophysically relevant parameters and Tables~\ref{tab:parConstraintsNoNus}-\ref{tab:parConstraintsAstroNu} report all fit parameter values. The parameters are defined as follows: $\gamma_\mathrm{inj}$ is the spectral index, $E^{\gamma_\mathrm{inj}}$, of the CRs injected into the source environment (i.e. the spectral index produced by the accelerator); $R_\mathrm{max}$ is the maximum rigidity of the injected CR spectrum, where the spectrum is cutoff exponentially; $r_\mathrm{esc}$ is the ratio of the escape-to-interaction time for the reference nucleus; $f_g$ is the fraction of interactions which are hadronic for the reference nucleus; $R_\mathrm{diff}$ is the rigidity scale of the magnetic field, assumed to be turbulent with a Kolmogorov spectrum; $r_\mathrm{size}$ is the ratio of the effective source size $L$ and the coherence length of the magnetic field $\lambda_c$; $f_\mathrm{gal}$ is the fraction of the observed flux at $10^{17.55}$~eV which is Galactic; $\gamma_\mathrm{gal}$ is the spectral index, $E^{\gamma_\mathrm{gal}}$, of the Galactic spectrum; $E^\mathrm{galFe}_\mathrm{max}$ is the maximum energy of Galactic iron, where the Galactic component is cutoff exponentially (this parameter sets the maximum rigidity of the Galactic component); $T$ is the black-body temperature of the photon spectral density distribution; $A_\mathrm{inj}$ is the mass number of the CRs injected into the source environment; $A_\mathrm{gal}$ is the mass number of the Galactic component (this component is also approximated as having a single mass); $B$ and $\lambda_c$ are the RMS strength and coherence length of the turbulent magnetic field; $n_\gamma = n_0 I_\mathrm{BB}(T)$ is the number density of photons; $n_\mathrm{g}$ is the number density of gas; and, $L$ is the effective size of the source environment. Parameter values for $n_0 < 1$ can be obtained from the black-body ($n_0 = 1$) values according to the following scalings: $L = L_\mathrm{BB}/n_0$, $B = B_\mathrm{BB} n_0$, $\lambda_c = \lambda_{c,\mathrm{BB}} / n_0$, and $n_\mathrm{g} = n_\mathrm{g,BB} n_0$. 

\begin{table}
\begin{tabular}{l c c}

\hline \hline
\textbf{Parameter} & \textbf{\textsc{Sibyll2.3c}} & \textbf{\textsc{EPOS-LHC}} \\
\hline
$\gamma_\mathrm{inj}$ & $-1.7^{+1.3}_{-0.93}$ & $-1.48^{+1.26}_{-1.01}$ \\
$\log_{10}(R_\mathrm{max}/\mathrm{V})$ & $18.69^{+0.79}_{-0.34}$ & $18.74^{+0.77}_{-0.34}$ \\
$\log_{10}{r_\mathrm{esc}}$ & $2.58^{+0.57}_{-0.75}$ & $2.38^{+0.61}_{-0.76}$ \\
$f_g$ & $0.04^{+0.16}_{-0.03}$ & $0.02^{+0.42}_{-0.02}$ \\
$\log_{10}(R_\mathrm{diff}/\mathrm{V})$ & $17.72^{+0.51}_{-0.94}$ & $17.7^{+0.57}_{-0.65}$ \\
$\tanh(\log_{10}{r_\mathrm{size}})$ & $0.95^{+0.04}_{-0.6}$ & $0.89^{+0.1}_{-0.55}$ \\
$f_\mathrm{gal}$ & $0.74^{+0.1}_{-0.33}$ & $0.75^{+0.08}_{-0.22}$ \\
$\gamma_\mathrm{gal}$ & $-3.44^{+0.45}_{-0.14}$ & $-3.5^{+0.26}_{-0.16}$ \\
$\log_{10}(E^\mathrm{galFe}_\mathrm{max}/\mathrm{eV})$ & $18.89^{+1.34}_{-0.47}$ & $18.63^{+0.61}_{-0.24}$ \\
$\log_{10}(T/\mathrm{K})$ & $2.68^{+0.97}_{-0.42}$ & $2.56^{+0.87}_{-0.43}$ \\
$A_\mathrm{inj}$ & $28.59^{+18.68}_{-18.82}$ & $28.45^{+18.74}_{-18.67}$ \\
$A_\mathrm{gal}$ & $28.62^{+18.55}_{-18.82}$ & $28.29^{+18.89}_{-18.57}$ \\
$\log_{10}(B\lambda_c/\mu\mathrm{G}\cdot\mathrm{kpc})$ & $0.56^{+0.51}_{-0.94}$ & $0.54^{+0.57}_{-0.65}$ \\
$\log_{10}(Ln_\gamma/(10\text{ kpc}\cdot\mathrm{cm}^{-3}))$ & $3.61^{+1.23}_{-1.2}$ & $3.86^{+0.8}_{-1.25}$ \\
$\log_{10}(n_\gamma/n_\mathrm{g})$ & $3.39^{+1.06}_{-0.68}$ & $3.65^{+1.47}_{-1.15}$ \\
$\log_{10}(L/10\text{ kpc})_\mathrm{BB}$ & $-6.45^{+2.53}_{-2.04}$ & $-5.85^{+2.29}_{-1.74}$ \\
$\log_{10}(\lambda_c/\mathrm{kpc})_\mathrm{BB}$ & $-7.67^{+4.07}_{-2.41}$ & $-6.45^{+3.1}_{-2.59}$ \\
$\log_{10}(n_\mathrm{g}/\mathrm{cm}^{-3})_\mathrm{BB}$ & $6.24^{+2.37}_{-1.44}$ & $5.71^{+1.61}_{-1.9}$ \\
$\log_{10}(B/\mu\mathrm{G})_\mathrm{BB}$ & $8.38^{+2.32}_{-4.66}$ & $7.18^{+2.45}_{-3.38}$ \\
\hline

\end{tabular}
\caption{\label{tab:parConstraintsNoNus}Preferred parameters (defined in the text) for the case fitting to the Auger spectrum and composition data~\citep{Verzi20,Yushkov20} alone for each HIM (we remind the reader, that this case is a not a valid fit). Central values denote the parameter median with uncertainties enclosing $68\%$ of the distribution about the median. Quantities labelled with subscript $\mathrm{BB}$ indicate quantities which rely on the assumption of a black-body ($n_0 = 1$) photon field; for other $n_0$ values $L = L_\mathrm{BB}/n_0$, $B = B_\mathrm{BB} n_0$, $\lambda_c = \lambda_{c,\mathrm{BB}} / n_0$, and $n_\mathrm{g} = n_\mathrm{g,BB} n_0$.}   

\end{table}

\begin{table}
\begin{tabular}{l c c}

\hline \hline
\textbf{Parameter} & \textbf{\textsc{Sibyll2.3c}} & \textbf{\textsc{EPOS-LHC}} \\
\hline
$\gamma_\mathrm{inj}$ & $-1.45^{+1.26}_{-1.21}$ & $-1.31^{+1.23}_{-1.17}$ \\
$\log_{10}(R_\mathrm{max}/\mathrm{V})$ & $18.63^{+0.81}_{-0.38}$ & $18.65^{+0.78}_{-0.37}$ \\
$\log_{10}{r_\mathrm{esc}}$ & $2.32^{+1.16}_{-0.92}$ & $2.01^{+1.49}_{-0.78}$ \\
$f_g$ & $0.17^{+0.6}_{-0.17}$ & $0.29^{+0.56}_{-0.29}$ \\
$\log_{10}(R_\mathrm{diff}/\mathrm{V})$ & $17.65^{+0.85}_{-1.7}$ & $17.7^{+1.01}_{-1.65}$ \\
$\tanh(\log_{10}{r_\mathrm{size}})$ & $0.81^{+0.18}_{-1.07}$ & $0.74^{+0.25}_{-1.02}$ \\
$f_\mathrm{gal}$ & $0.71^{+0.16}_{-0.47}$ & $0.76^{+0.08}_{-0.49}$ \\
$\gamma_\mathrm{gal}$ & $-3.4^{+0.74}_{-0.21}$ & $-3.46^{+0.74}_{-0.23}$ \\
$\log_{10}(E^\mathrm{galFe}_\mathrm{max}/\mathrm{eV})$ & $18.86^{+1.35}_{-0.63}$ & $18.66^{+1.45}_{-0.47}$ \\
$\log_{10}(T/\mathrm{K})$ & $2.41^{+0.85}_{-0.6}$ & $2.21^{+1.05}_{-0.39}$ \\
$A_\mathrm{inj}$ & $28.83^{+18.78}_{-18.83}$ & $28.62^{+18.93}_{-18.71}$ \\
$A_\mathrm{gal}$ & $28.78^{+18.77}_{-18.8}$ & $28.7^{+18.8}_{-18.72}$ \\
$\log_{10}(B\lambda_c/\mu\mathrm{G}\cdot\mathrm{kpc})$ & $0.49^{+0.85}_{-1.7}$ & $0.54^{+1.01}_{-1.65}$ \\
$\log_{10}(Ln_\gamma/(10\text{ kpc}\cdot\mathrm{cm}^{-3}))$ & $3.96^{+3.09}_{-1.51}$ & $4.15^{+2.65}_{-1.48}$ \\
$\log_{10}(n_\gamma/n_\mathrm{g})$ & $3.17^{+1.7}_{-1.18}$ & $3.05^{+2.06}_{-1.22}$ \\
$\log_{10}(L/10\text{ kpc})_\mathrm{BB}$ & $-4.7^{+2.75}_{-2.4}$ & $-4.12^{+1.87}_{-2.56}$ \\
$\log_{10}(\lambda_c/\mathrm{kpc})_\mathrm{BB}$ & $-4.71^{+3.91}_{-4.0}$ & $-4.23^{+3.11}_{-3.2}$ \\
$\log_{10}(n_\mathrm{g}/\mathrm{cm}^{-3})_\mathrm{BB}$ & $5.52^{+2.25}_{-2.17}$ & $5.08^{+3.2}_{-2.24}$ \\
$\log_{10}(B/\mu\mathrm{G})_\mathrm{BB}$ & $5.19^{+4.33}_{-4.97}$ & $4.64^{+3.77}_{-4.02}$ \\
\hline

\end{tabular}
\caption{\label{tab:parConstraints}Same as Table~\ref{tab:parConstraintsNoNus} for the case fitting to the Auger spectrum and composition data~\citep{Verzi20,Yushkov20} and compatible with IceCube bounds on neutrinos above $10^{15.9}$~eV~\citep{Aartsen+18} for each HIM.}   

\end{table}

\begin{table}
\begin{tabular}{l c c}

\hline \hline
\textbf{Parameter} & \textbf{\textsc{Sibyll2.3c}} & \textbf{\textsc{EPOS-LHC}} \\
\hline
$\gamma_\mathrm{inj}$ & $-1.34^{+1.19}_{-0.96}$ & $-1.23^{+1.16}_{-0.99}$ \\
$\log_{10}(R_\mathrm{max}/\mathrm{V})$ & $18.64^{+0.62}_{-0.36}$ & $18.67^{+0.66}_{-0.38}$ \\
$\log_{10}{r_\mathrm{esc}}$ & $2.23^{+1.22}_{-0.88}$ & $2.03^{+1.41}_{-0.8}$ \\
$f_g$ & $0.09^{+0.63}_{-0.08}$ & $0.17^{+0.57}_{-0.17}$ \\
$\log_{10}(R_\mathrm{diff}/\mathrm{V})$ & $17.71^{+0.84}_{-1.83}$ & $17.73^{+0.99}_{-1.71}$ \\
$\tanh(\log_{10}{r_\mathrm{size}})$ & $0.75^{+0.24}_{-1.12}$ & $0.71^{+0.27}_{-1.09}$ \\
$f_\mathrm{gal}$ & $0.73^{+0.14}_{-0.46}$ & $0.77^{+0.07}_{-0.48}$ \\
$\gamma_\mathrm{gal}$ & $-3.44^{+0.7}_{-0.19}$ & $-3.48^{+0.74}_{-0.23}$ \\
$\log_{10}(E^\mathrm{galFe}_\mathrm{max}/\mathrm{eV})$ & $18.83^{+1.27}_{-0.56}$ & $18.63^{+1.33}_{-0.43}$ \\
$\log_{10}(T/\mathrm{K})$ & $2.38^{+0.74}_{-0.53}$ & $2.23^{+0.98}_{-0.37}$ \\
$A_\mathrm{inj}$ & $28.62^{+18.75}_{-18.91}$ & $28.84^{+18.89}_{-18.93}$ \\
$A_\mathrm{gal}$ & $28.53^{+18.88}_{-18.76}$ & $28.47^{+19.17}_{-18.8}$ \\
$\log_{10}(B\lambda_c/\mu\mathrm{G}\cdot\mathrm{kpc})$ & $0.55^{+0.84}_{-1.83}$ & $0.57^{+0.99}_{-1.71}$ \\
$\log_{10}(Ln_\gamma/(10\text{ kpc}\cdot\mathrm{cm}^{-3}))$ & $4.0^{+2.88}_{-1.41}$ & $4.14^{+2.5}_{-1.03}$ \\
$\log_{10}(n_\gamma/n_\mathrm{g})$ & $3.33^{+1.46}_{-1.19}$ & $3.44^{+1.71}_{-1.37}$ \\
$\log_{10}(L/10\text{ kpc})_\mathrm{BB}$ & $-4.54^{+2.21}_{-2.19}$ & $-3.97^{+1.56}_{-2.41}$ \\
$\log_{10}(\lambda_c/\mathrm{kpc})_\mathrm{BB}$ & $-4.45^{+3.45}_{-3.63}$ & $-4.06^{+2.74}_{-2.71}$ \\
$\log_{10}(n_\mathrm{g}/\mathrm{cm}^{-3})_\mathrm{BB}$ & $5.29^{+2.1}_{-1.83}$ & $4.89^{+2.59}_{-2.0}$ \\
$\log_{10}(B/\mu\mathrm{G})_\mathrm{BB}$ & $4.96^{+4.04}_{-4.71}$ & $4.5^{+3.31}_{-3.77}$ \\
\hline

\end{tabular}
\caption{\label{tab:parConstraintsAstroNu}Same as Table~\ref{tab:parConstraintsNoNus} for the case fitting to both the Auger spectrum and composition data and~\citep{Verzi20,Yushkov20} and the IceCube astrophysical neutrino data~\citep{IceCubeCascades20,IceCubeGlashow2021}, while being compatible with IceCube bounds on neutrinos above $10^{15.9}$~eV~\citep{Aartsen+18} for each HIM.}   

\end{table}

\section{Black-body $B$ vs $L$: Joint posterior distributions and astrophysical sources} 

\par
Figure~\ref{fig:BL_constraints} shows the joint posterior distribution between $B$ and $L$ for $n_0=1$, for both HIMs and three analysis cases.
As a reminder, results fitting to CR data alone (Figs.~\ref{fig:BL_constraints_CR_noNus_sibyll} and~\ref{fig:BL_constraints_CR_noNus_epos}) are presented mostly as an illustrative exercise to show the impact of including EHE neutrino bounds in the analysis; it is not possible to draw conclusions from the case fitting UHECR data alone as the EHE neutrino bounds must be respected. The two physical cases show a remarkable consistency, owing to the relatively strong constraints presented by the combination of UHECR data and EHE neutrino bounds. The addition of simultaneously fitting to astrophysical neutrino data only slightly shrinks the $1\sigma$ region. In all three analysis cases, the results are largely independent of the HIM assumed.

\par
When attempting to draw conclusions from the plots in Fig.~\ref{fig:BL_constraints} it is important to keep in mind that 
they are for a black-body-like source environment, $n_0=1$, which may not be applicable. However, using the scalings given in the previous section one can see that the effect of a grey-body-like source environment (i.e. $n_0 < 1$) is to shift the joint posterior distribution (grey regions) toward higher values of $L$ and lower values of $B$ by a factor of $n_0$ (i.e. the distribution shifts downwards along the black ``Hillas rails''). Note that, given our definition, $n_0$  
is the emissivity of the source environment and must, therefore, satisfy $0 \leq n_0 \leq 1$ if the photon field is in fact grey-body. For self-consistency, one must also verify that the typical emissivity of a given candidate source type is compatible with the chosen value of $n_0$. For $n_0$-independent results the reader is referred to the following section.

\par
The size of the $3\sigma$ region in Fig.~\ref{fig:BL_constraints} is markedly smaller for the case where only UHECR data is considered (top row panels). This illustrates that UHECR data alone is extremely constraining, and also shows that studies omitting neutrino bounds are likely to overestimate their ability to constrain astrophysical parameters. In other words, the posterior distribution for the UHECR-only case is highly peaked for many parameters, but many of those points in parameter space also violate EHE neutrino bounds. Enforcing the neutrino bounds reduces the peakedness of the posterior distributions, resulting in a broader distribution and less constrained parameters.

\par
In addition to the main peak in the joint posterior distribution, at $L \lesssim 100$~pc and $B \gtrsim 0.1$~mG, there is a less significant peak at $L \gtrsim 1$~Mpc and $B \lesssim 1$ $\mu$G, consistent across all analysis cases and HIMs. 

\section{Possible significance of synchrotron cooling}\label{app:syncCooling}

\par
\begin{figure*}[htpb!]
	\centering
	\begin{minipage}{0.49\linewidth}
	  \centering
      \subfloat[\label{fig:BL_constraints_CR_noNus_sibyll}]{\includegraphics[width=\textwidth]{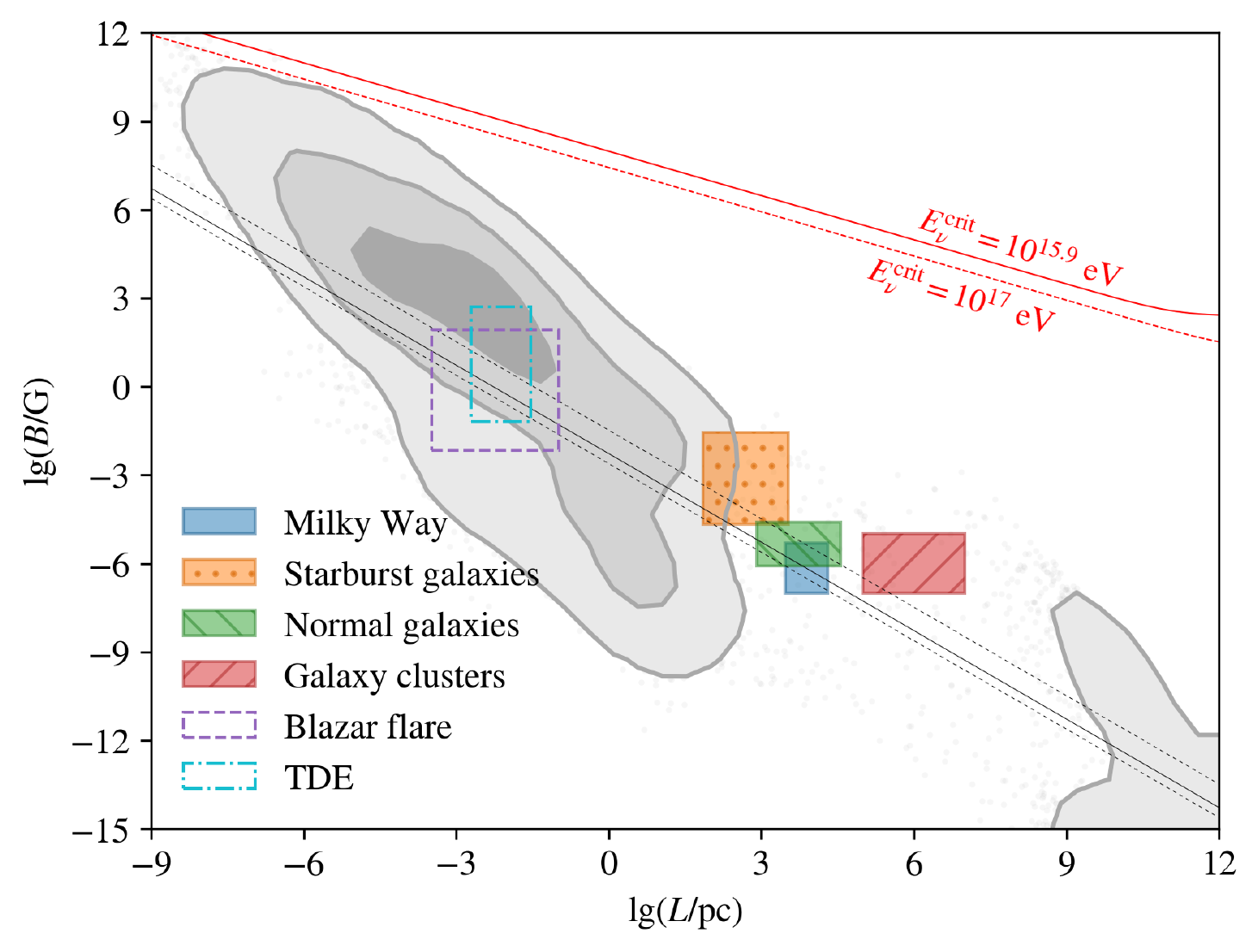}}
    \end{minipage}
    \begin{minipage}{0.49\linewidth}
	  \centering
      \subfloat[\label{fig:BL_constraints_CR_noNus_epos}]{\includegraphics[width=\textwidth]{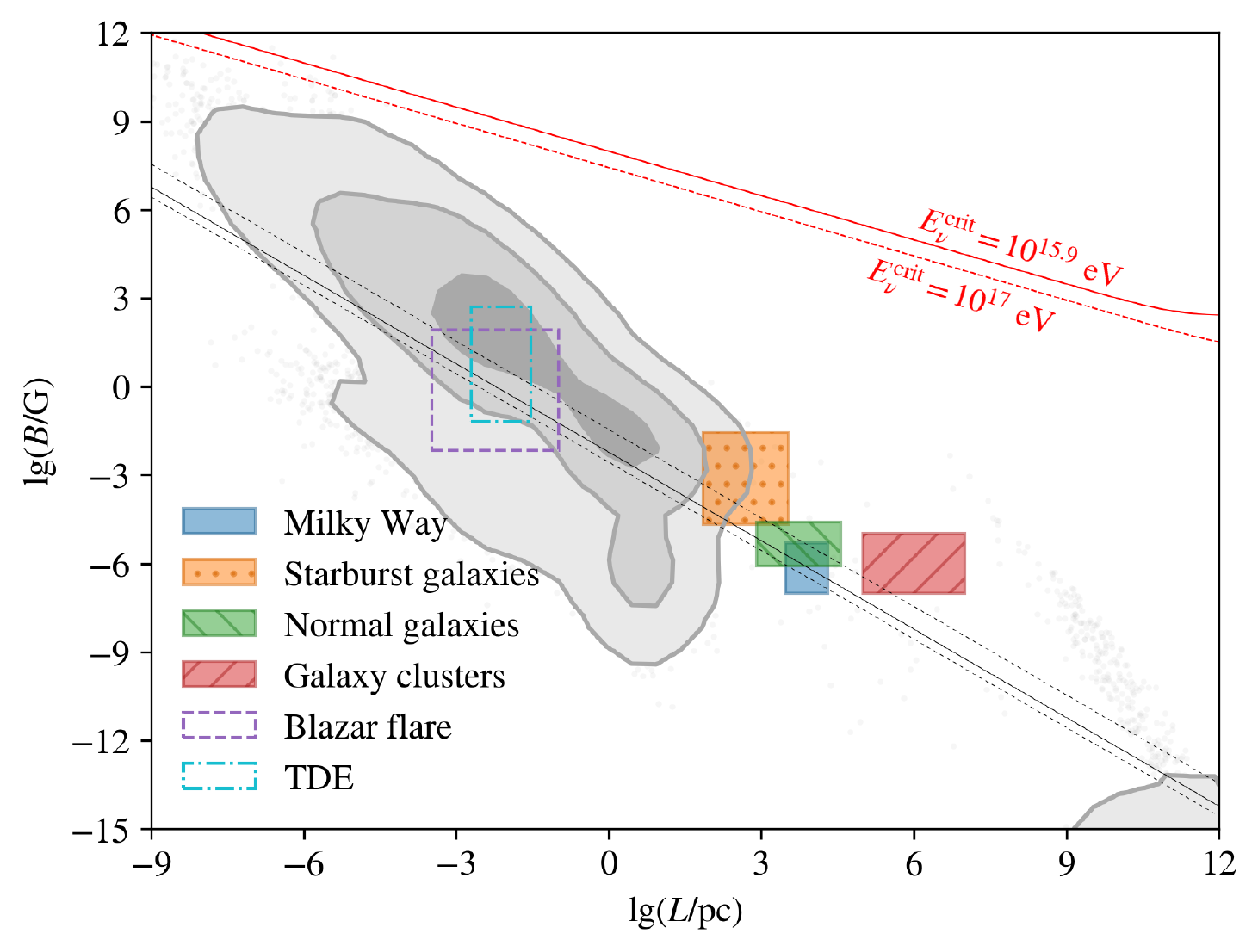}}
    \end{minipage}
    \begin{minipage}{0.49\linewidth}
	  \centering
      \subfloat[\label{fig:BL_constraints_CR_sibyll}]{\includegraphics[width=\textwidth]{bestShifts_sibyll_BB_limDiffusion_BL_constraints.pdf}}
    \end{minipage}
    \begin{minipage}{0.49\linewidth}
	  \centering
      \subfloat[\label{fig:BL_constraints_CR_epos}]{\includegraphics[width=\textwidth]{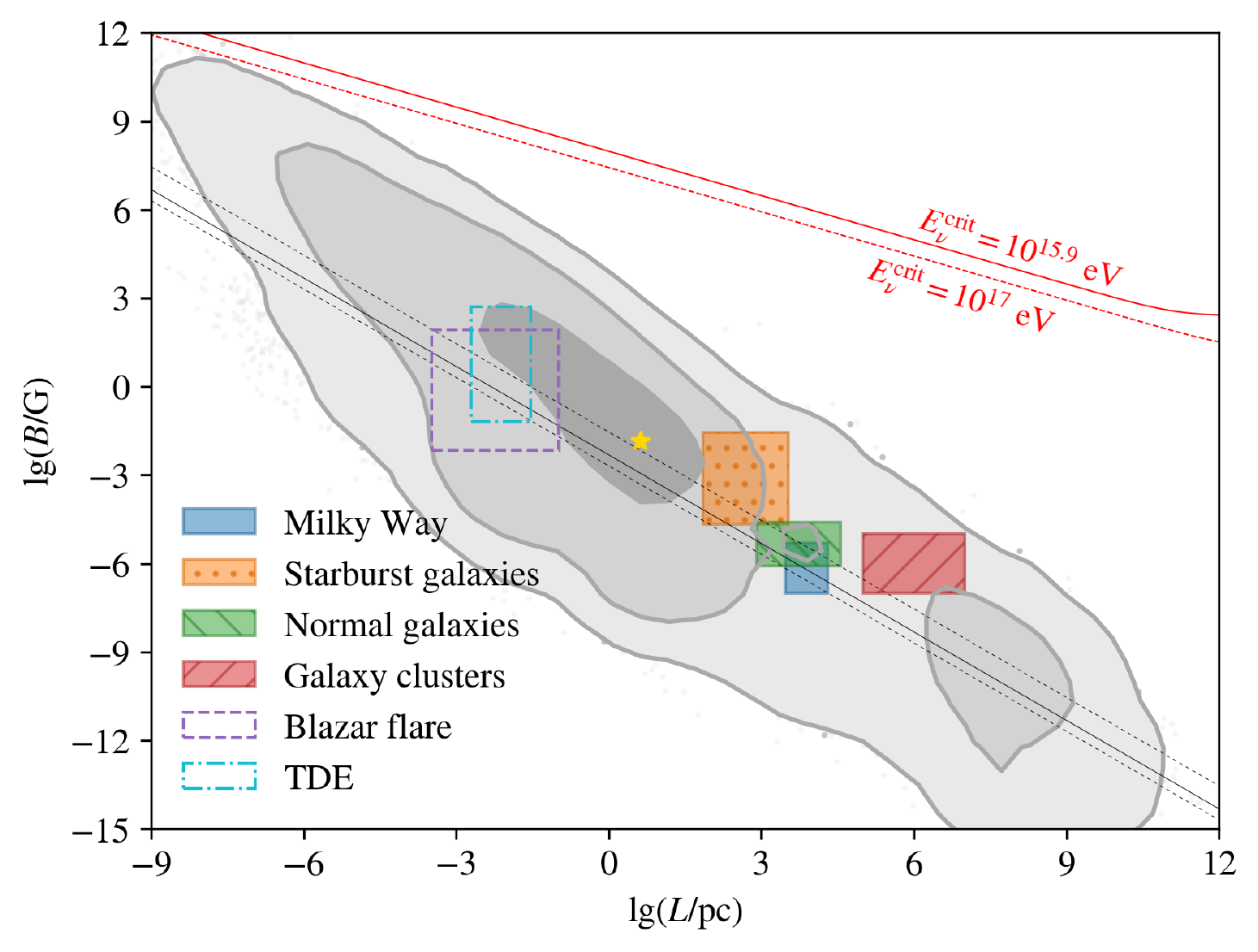}}
    \end{minipage}
    \begin{minipage}{0.49\linewidth}
	  \centering
      \subfloat[\label{fig:BL_constraints_CR_astroNu_sibyll}]{\includegraphics[width=\textwidth]{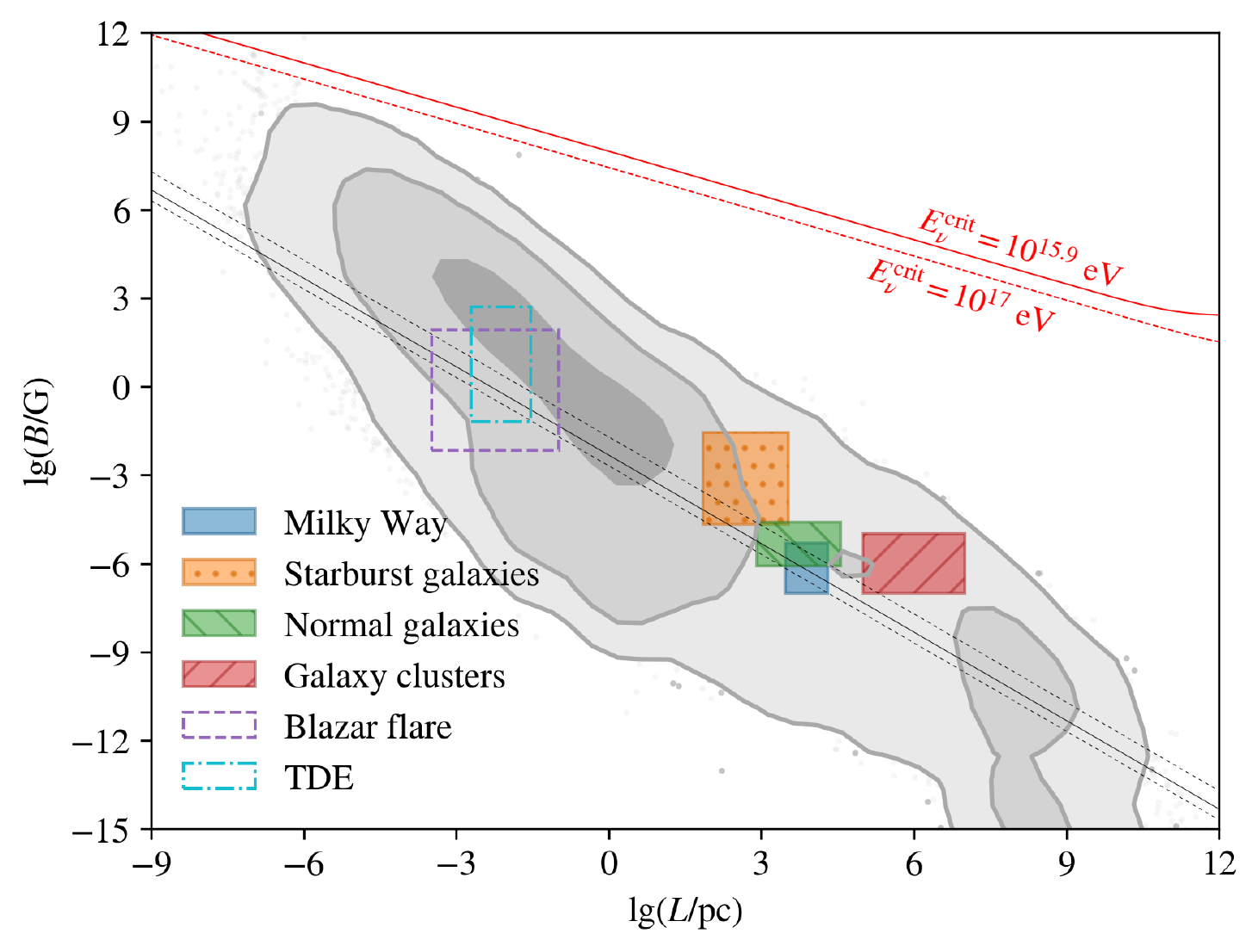}}
    \end{minipage}
    \begin{minipage}{0.49\linewidth}
	  \centering
      \subfloat[\label{fig:BL_constraints_CR_astroNu_epos}]{\includegraphics[width=\textwidth]{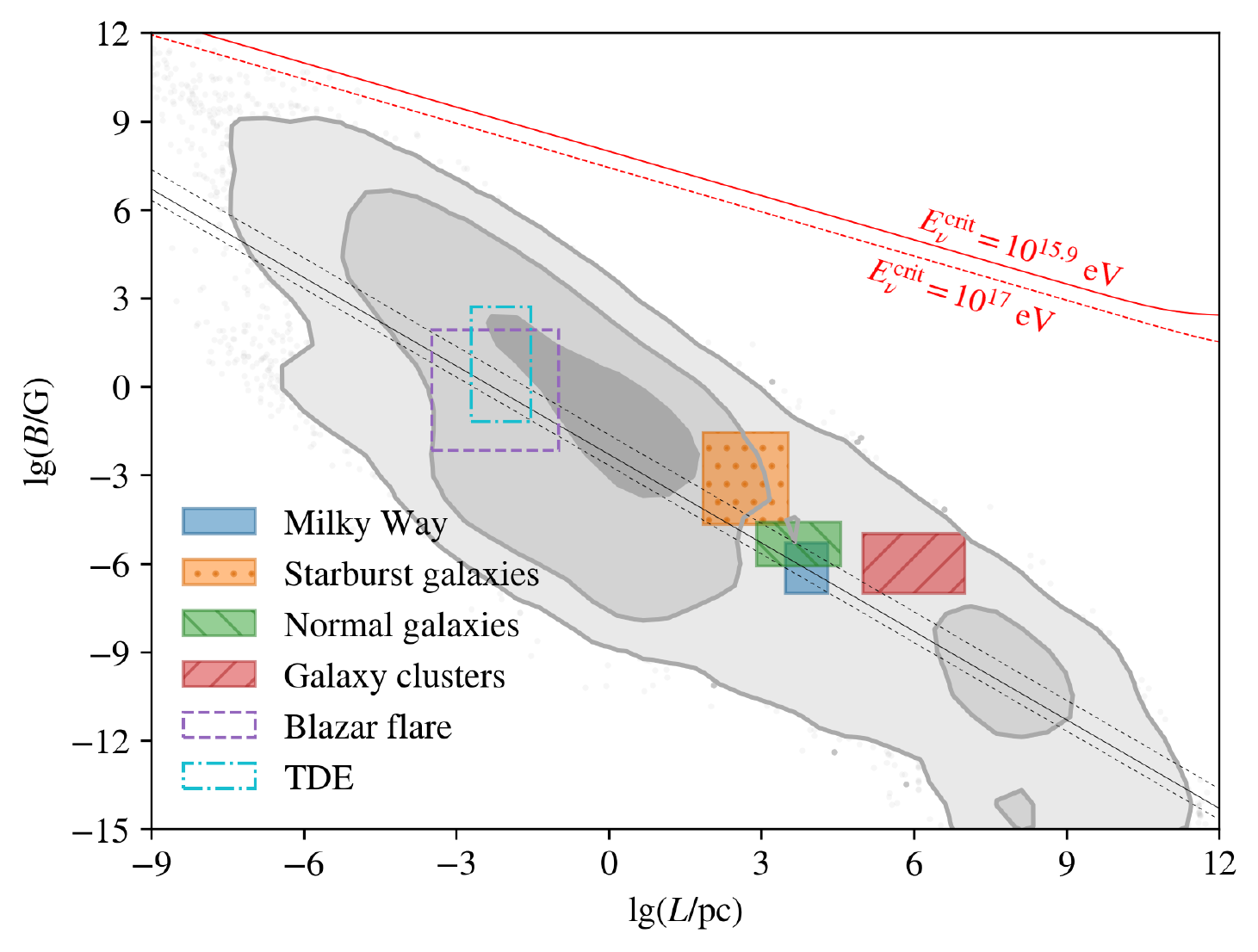}}
    \end{minipage}
	\caption{The joint posterior distribution of the effective size, $L$, and magnetic field strength, $B$, of the source environment fitting to the UHECR spectrum and composition of Auger alone (top), while compatible with IceCube bounds on EHE neutrinos (middle), and while simultaneously fitting to astrophysical neutrino data (bottom), using \textsc{Sibyll2.3c} (left) or \textsc{EPOS-LHC} (right), and taking a black-body spectrum, $n_0 = 1$. The case fitting UHECR data alone (top row) is not a valid fit. The bands give the $1\sigma$, $2\sigma$, and $3\sigma$ uncertainty bands (darkest to lightest grey, respectively) of the joint posterior distribution. For a different value of $n_0$, the posterior distribution slides along the diagonal as discussed in the text. The maximum rigidity of the \textit{accelerator} is shown for the median and $16$th/$84$th percentiles (solid and dashed black lines, respectively) of the posterior distribution for $\log_{10}R_\mathrm{max}$. Red lines demarcate regions where synchrotron losses in the source environment significantly affect the neutrino spectrum and a more detailed analysis would be required. The indicated size and magnetic field strengths of various potential source types are approximate and serve as a guide.}
	\label{fig:BL_constraints}
\end{figure*}

The posterior distributions shown in Figs.~\ref{fig:BL_constraints_CR_sibyll}-\ref{fig:BL_constraints_CR_astroNu_epos} have excluded models which violate bounds on EHE neutrinos. In principle, these bounds could be evaded if the charged pions and muons producing the neutrinos suffer significant synchrotron losses in the source environment before escaping or decaying. However as we now show, this is not the case. To check whether this applies to our analysis we calculate the curve in the $B-L$ plane above which the effects of synchrotron losses are significant for neutrinos beyond a critical energy, $E^\mathrm{crit}_\nu$. We obtain this curve by equating the synchrotron loss time for a $3 E^\mathrm{crit}_\nu$ muon to the harmonic sum of its decay and escape times, for a given $L$ and $B$. (The synchrotron loss time depends on $\lambda_c$ as well, through the muon's escape time, but in practice this dependence is weak.) The results are plotted in solid and dashed red lines in Fig.~\ref{fig:BL_constraints}; below these curves synchrotron losses are insignificant for the neutrino spectrum below $E^{\rm crit}_\nu = 10^{15.9}$~eV and $10^{17}$~eV, respectively. Since our models only produce a significant neutrino flux up to at most $\sim 10^{17}$~eV (see~MFU22) and the joint posterior distribution obtained from our analysis lies below the boundary of the region in which cooling is important for $E^\mathrm{crit}_\nu = 10^{17}$~eV, performing the fits including neutrino bounds but ignoring cooling is self-consistent. This is true even when only CR data is considered, irrespective of neutrino bounds as is seen in Figs.~\ref{fig:BL_constraints_CR_noNus_sibyll} and~\ref{fig:BL_constraints_CR_noNus_epos}.\\

\section{$BL$ vs \texorpdfstring{\textit{L\MakeLowercase{n}}\MakeLowercase{\textsubscript{g}}}: Joint posterior distributions and astrophysical sources}

\begin{figure*}[htpb!]
	\centering
	\begin{minipage}{0.49\linewidth}
	  \centering
      \subfloat[\label{fig:BL_LnGas_constraints_CR_noNus_sibyll}]{\includegraphics[width=\textwidth]{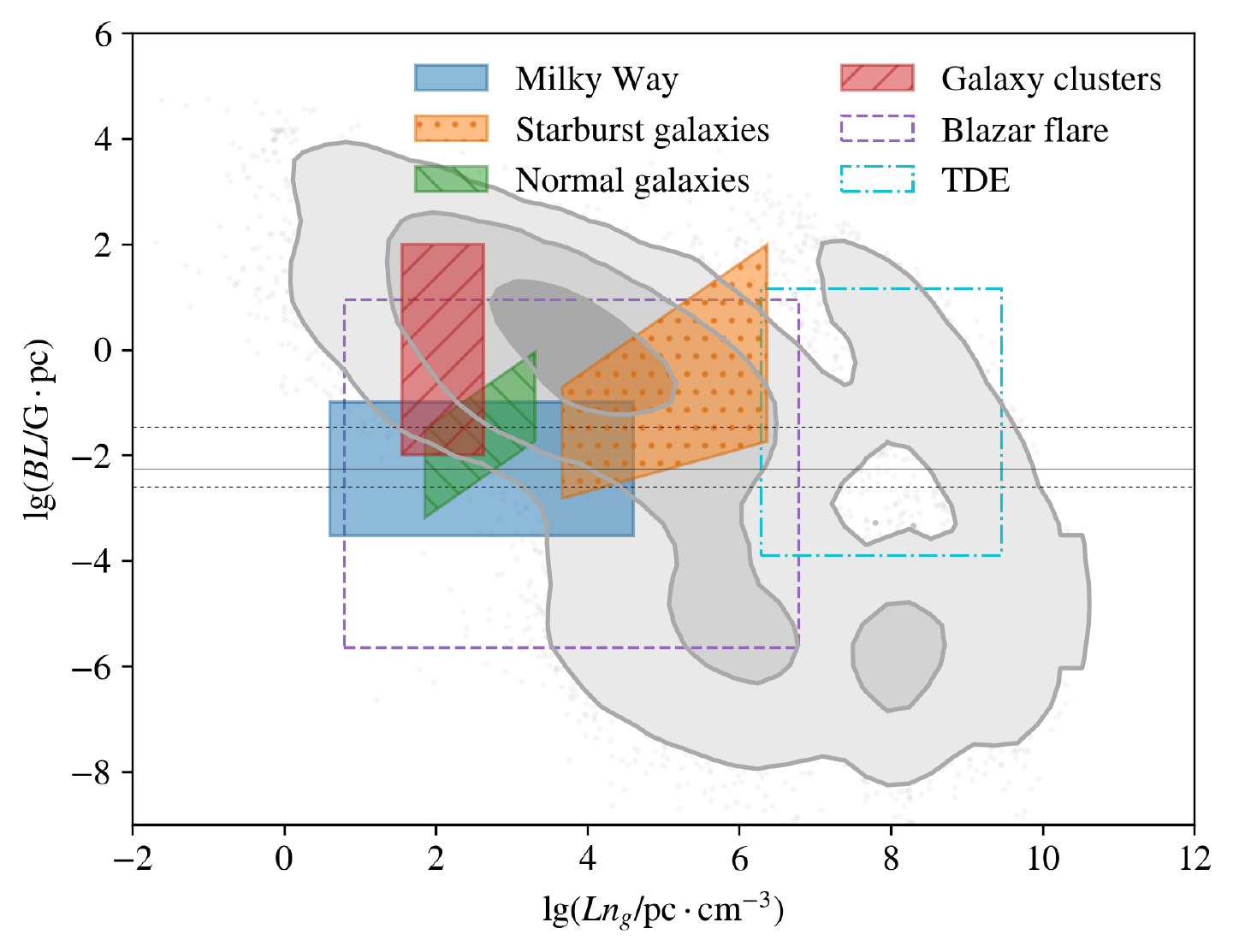}}
    \end{minipage}
    \begin{minipage}{0.49\linewidth}
	  \centering
      \subfloat[\label{fig:BL_LnGas_constraints_CR_noNus_epos}]{\includegraphics[width=\textwidth]{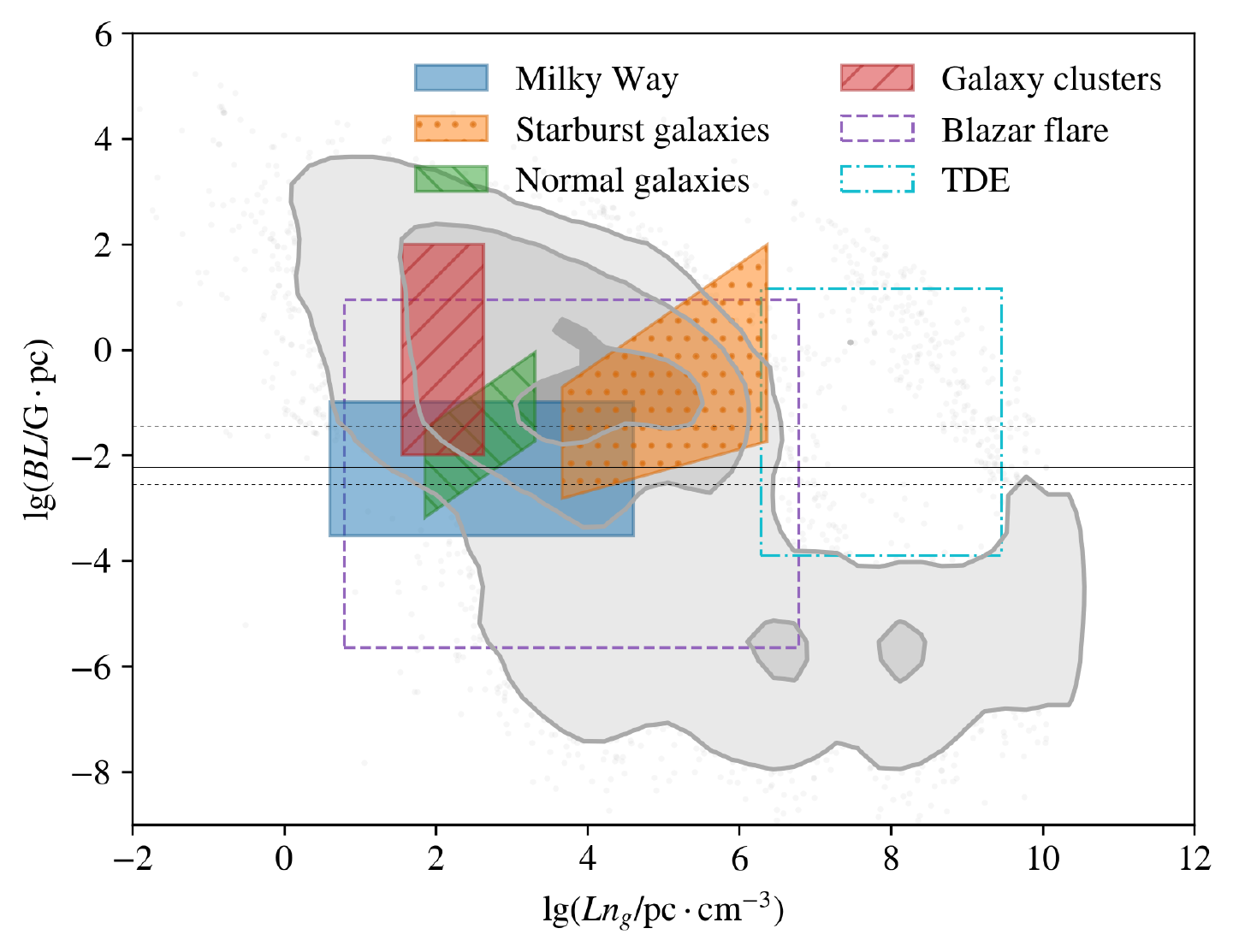}}
    \end{minipage}
    \begin{minipage}{0.49\linewidth}
	  \centering
      \subfloat[\label{fig:BL_LnGas_constraints_CR_sibyll}]{\includegraphics[width=\textwidth]{bestShifts_sibyll_BB_limDiffusion_BL_LnGas_constraints.pdf}}
    \end{minipage}
    \begin{minipage}{0.49\linewidth}
	  \centering
      \subfloat[\label{fig:BL_LnGas_constraints_CR_epos}]{\includegraphics[width=\textwidth]{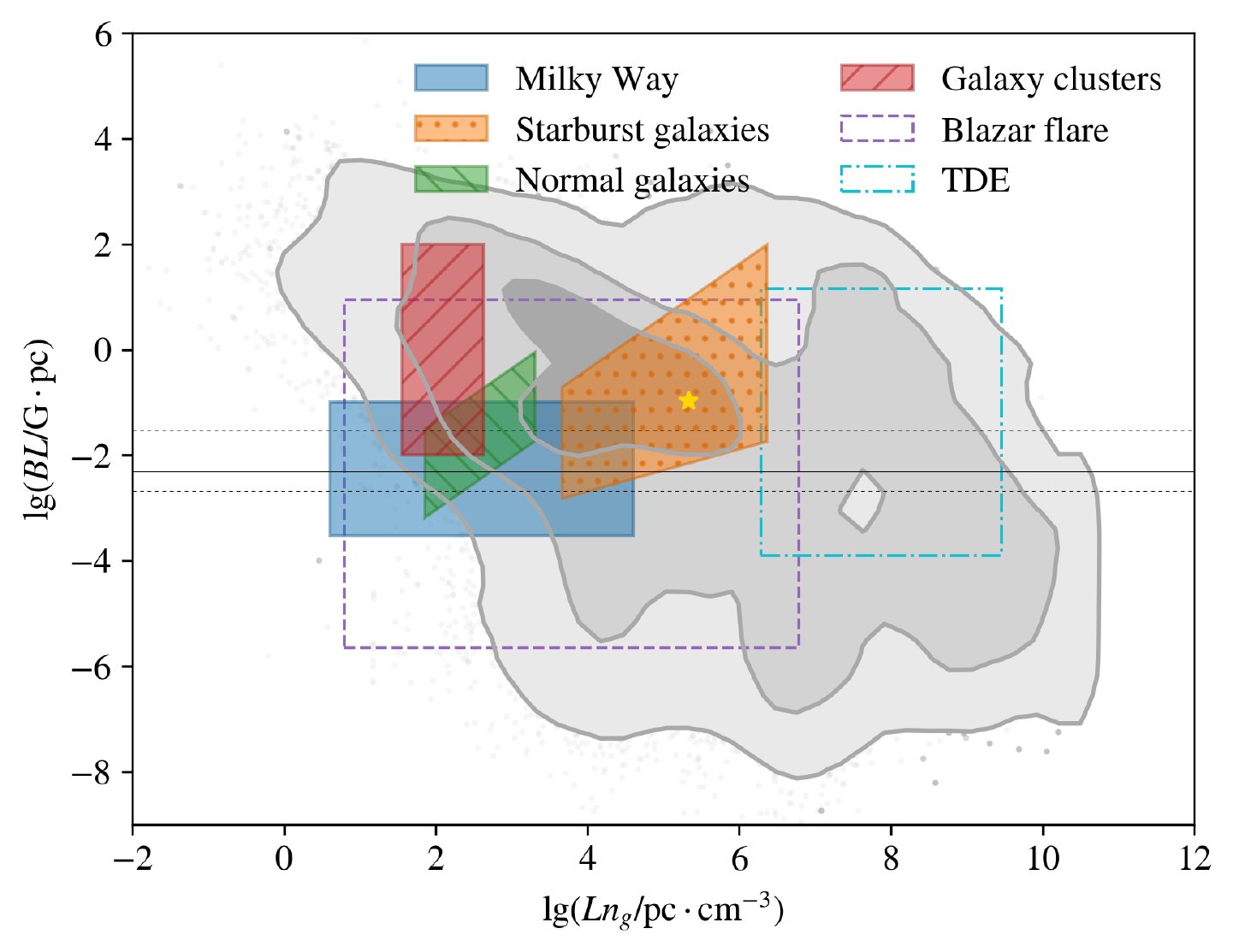}}
    \end{minipage}
    \begin{minipage}{0.49\linewidth}
	  \centering
      \subfloat[\label{fig:BL_LnGas_constraints_CR_astroNu_sibyll}]{\includegraphics[width=\textwidth]{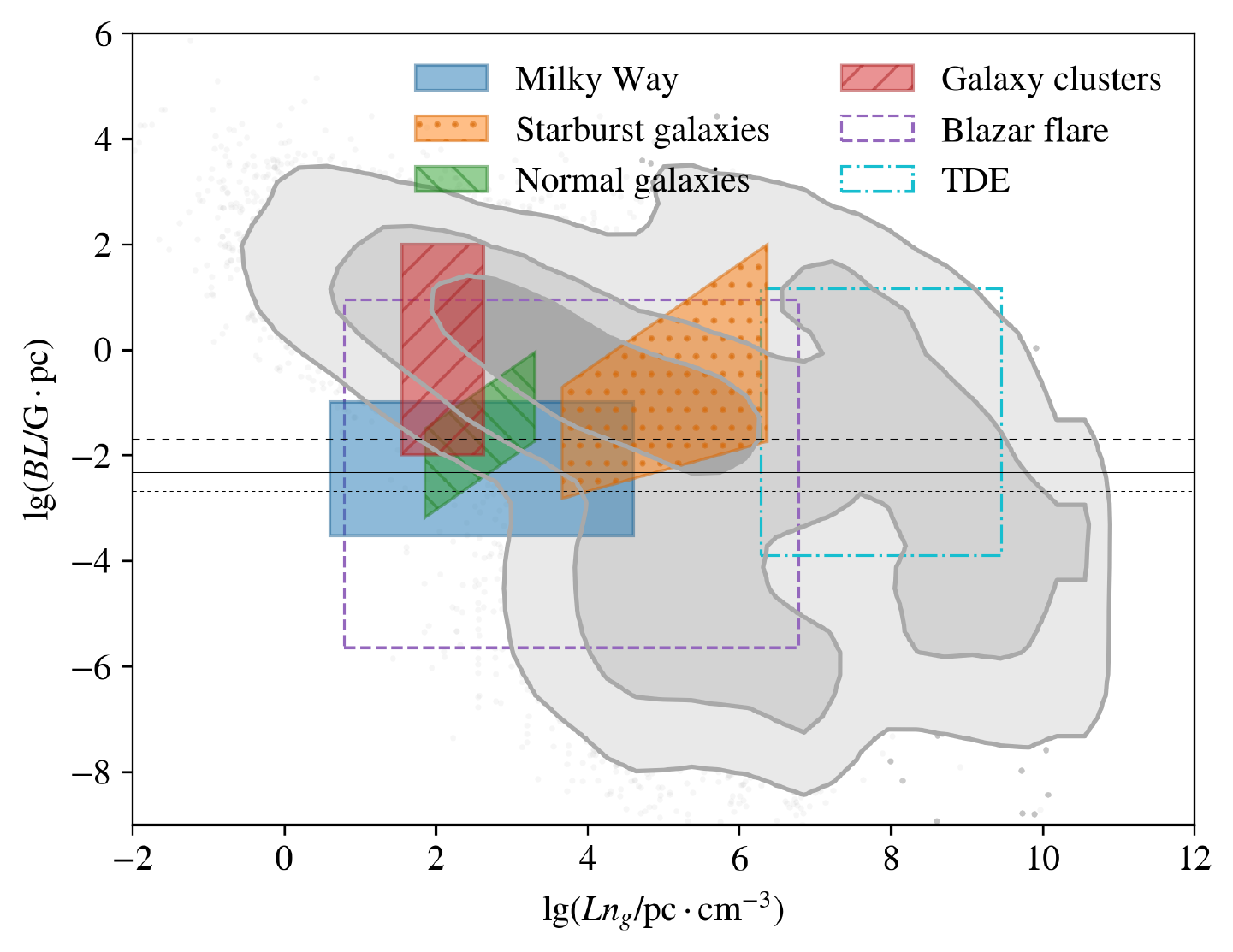}}
    \end{minipage}
    \begin{minipage}{0.49\linewidth}
	  \centering
      \subfloat[\label{fig:BL_LnGas_constraints_CR_astroNu_epos}]{\includegraphics[width=\textwidth]{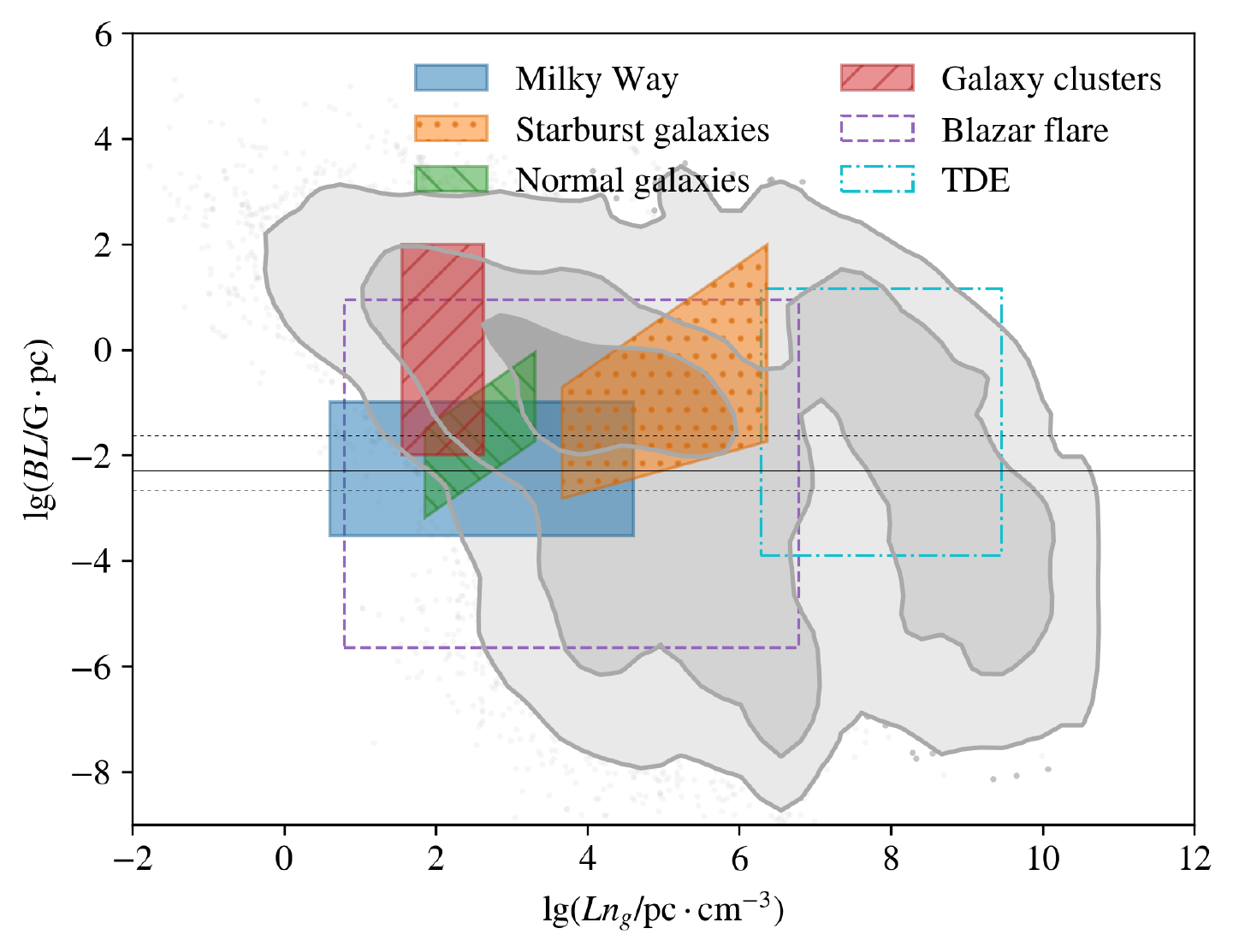}}
    \end{minipage}
	\caption{Same as Fig.~\ref{fig:BL_constraints} for $B \, L$ and $L n_\mathrm{g}$. These products are independent of the value of $n_0$ so that the joint posterior distribution is unaffected by its value.}
	\label{fig:BL_LnGas_constraints}
\end{figure*}

\par
Figure~\ref{fig:BL_LnGas_constraints} shows the joint posterior distribution between $BL$ and $Ln_\mathrm{g}$ for both HIMs and three analysis cases. 
The motivation for these plots is that they are independent of the value of $n_0$. 
As in the previous section, a UHECR-only analysis (Figs.~\ref{fig:BL_LnGas_constraints_CR_noNus_sibyll} and~\ref{fig:BL_LnGas_constraints_CR_noNus_epos}) results in stronger constraints than one considering bounds on EHE neutrinos. We emphasize, again, that the UHECR-only case is just for illustrative purposes and that conclusions about UHECR sources cannot be drawn from an analysis that neglects neutrino constraints. 

\par
The results in Fig.~\ref{fig:BL_LnGas_constraints} show a remarkable consistency, irrespective of the assumed HIM, favoring surface number densities $\Sigma_\mathrm{g}/m_p \simeq Ln_\mathrm{g}$ between $\sim 10^2$ and $\sim 10^6$~pc/cm$^3$ and $BL \gtrsim 10^{-3}$~G$\cdot$pc. These results would seem to favor source types like starburst galaxies (SBGs) and active galactic nuclei (AGN).

\section{Astrophysical parameter corner plots}\label{app:cornerplots}

\par
Figures~\ref{fig:astroEnv_CR_sibyll} and \ref{fig:astroEnv_CR_epos} show corner plots for some important astrophysical parameters for each HIM in our fiducial model -- fitting to UHECR data alone while remaining compatible with IceCube bounds on EHE neutrinos. These results assume a black-body-like source environment ($n_0 = 1$), but the corresponding results for grey-body-like source environments ($n_0 < 1$) can be obtained according to the scalings given in Appendix~\ref{app:preferredparams}. Note that $T$ does not scale with $n_0$ as it is directly a fit parameter. These corner plots and posterior distributions serve as an additional set of criteria which environments of candidate UHECR sources must satisfy in order to be compatible with current UHECR data and neutrino bounds.

\begin{figure*}[htpb!]
	\centering
    \includegraphics[width=\textwidth]{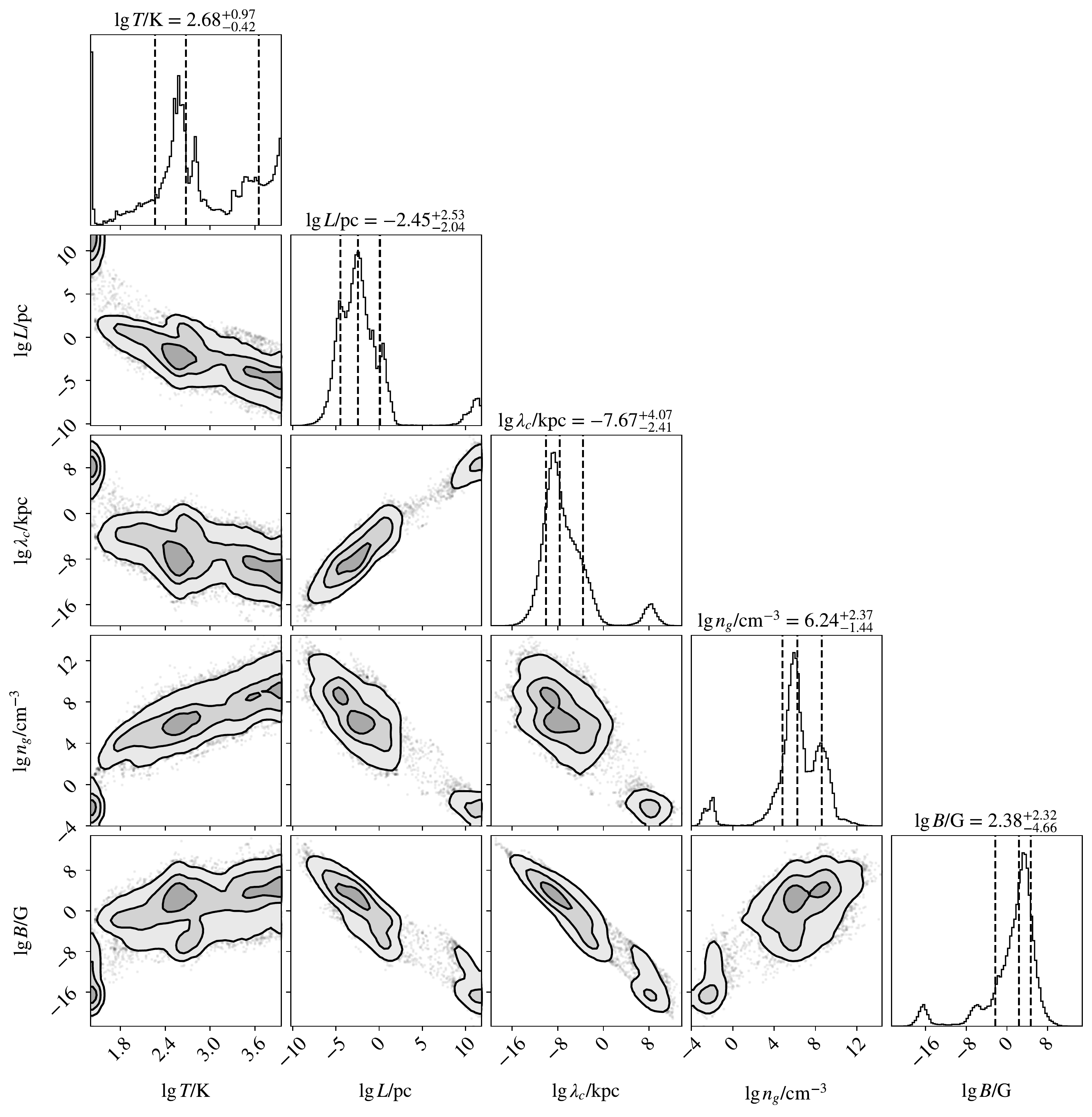}
    \caption{Posterior distribution of astrophysical parameters for our fidicual model, fitting to UHECR data alone using the \textsc{Sibyll2.3c} HIM. Dotted lines on one-dimensional histograms indicate the median, and $16$th and $84$th percentiles of the distribution. Gray regions on two-dimensional histograms denote the $1\sigma$, $2\sigma$, and  $3\sigma$ uncertainty bands of the distribution (darkest to lightest, respectively). 
	These plots are for $n_0 = 1$, a black-body photon spectrum.}
    \label{fig:astroEnv_CR_only_sibyll}
\end{figure*}

\begin{figure*}[htpb!]
	\centering
    \includegraphics[width=\textwidth]{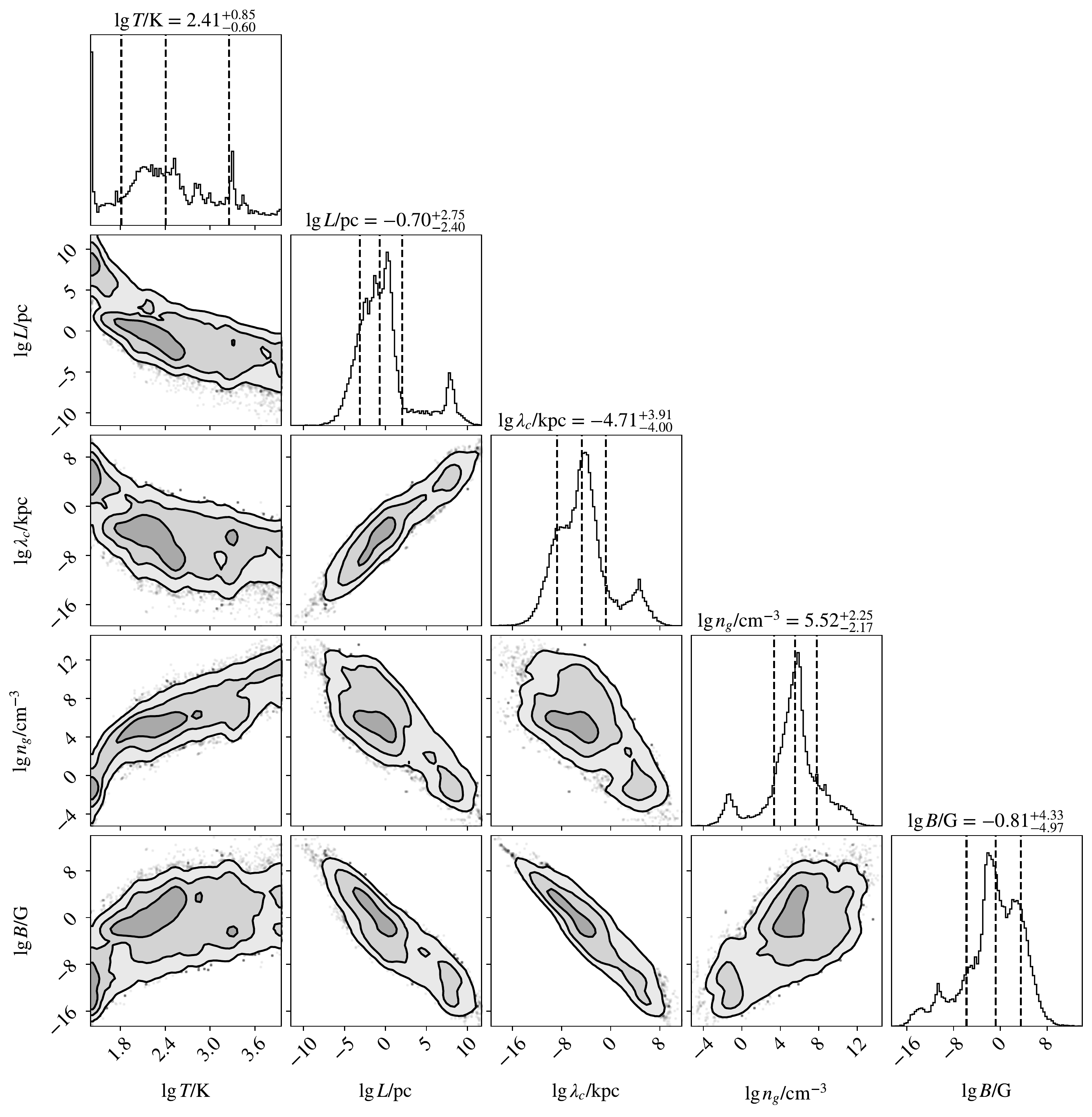}
    \caption{Same as Fig.~\ref{fig:astroEnv_CR_only_sibyll} but fitting to UHECR data alone while remaining compatible with IceCube bounds on EHE neutrinos under the assumption of the \textsc{Sibyll2.3c} HIM.}
    \label{fig:astroEnv_CR_sibyll}
\end{figure*}

\begin{figure*}[htpb!]
    \includegraphics[width=\textwidth]{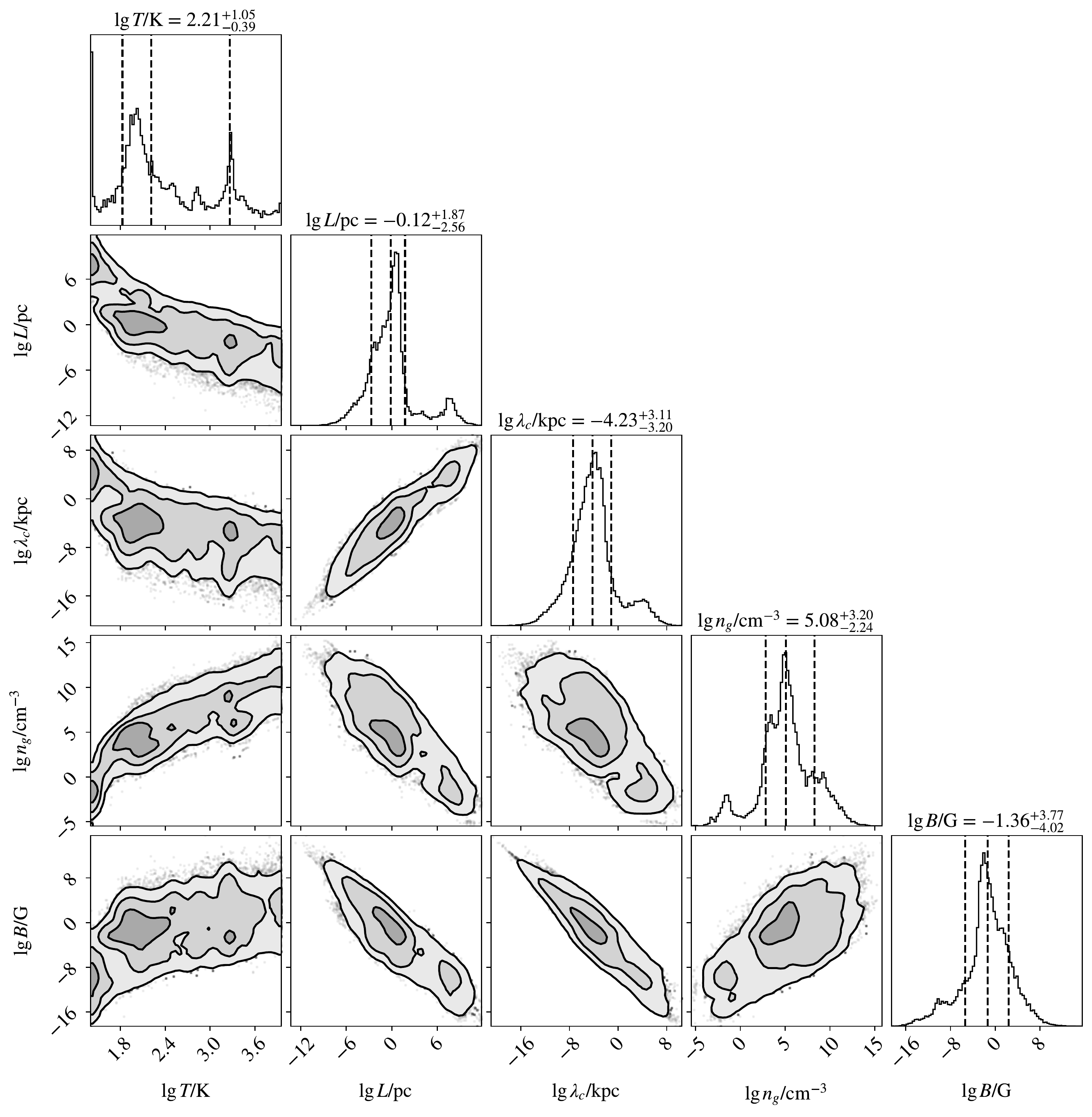}
	\caption{Same as Fig.~\ref{fig:astroEnv_CR_sibyll} for the \textsc{EPOS-LHC} HIM.}
    \label{fig:astroEnv_CR_epos}
\end{figure*}

\section{Effect of source evolution}

\par
\begin{figure}
	\centering
    \includegraphics[width=\linewidth]{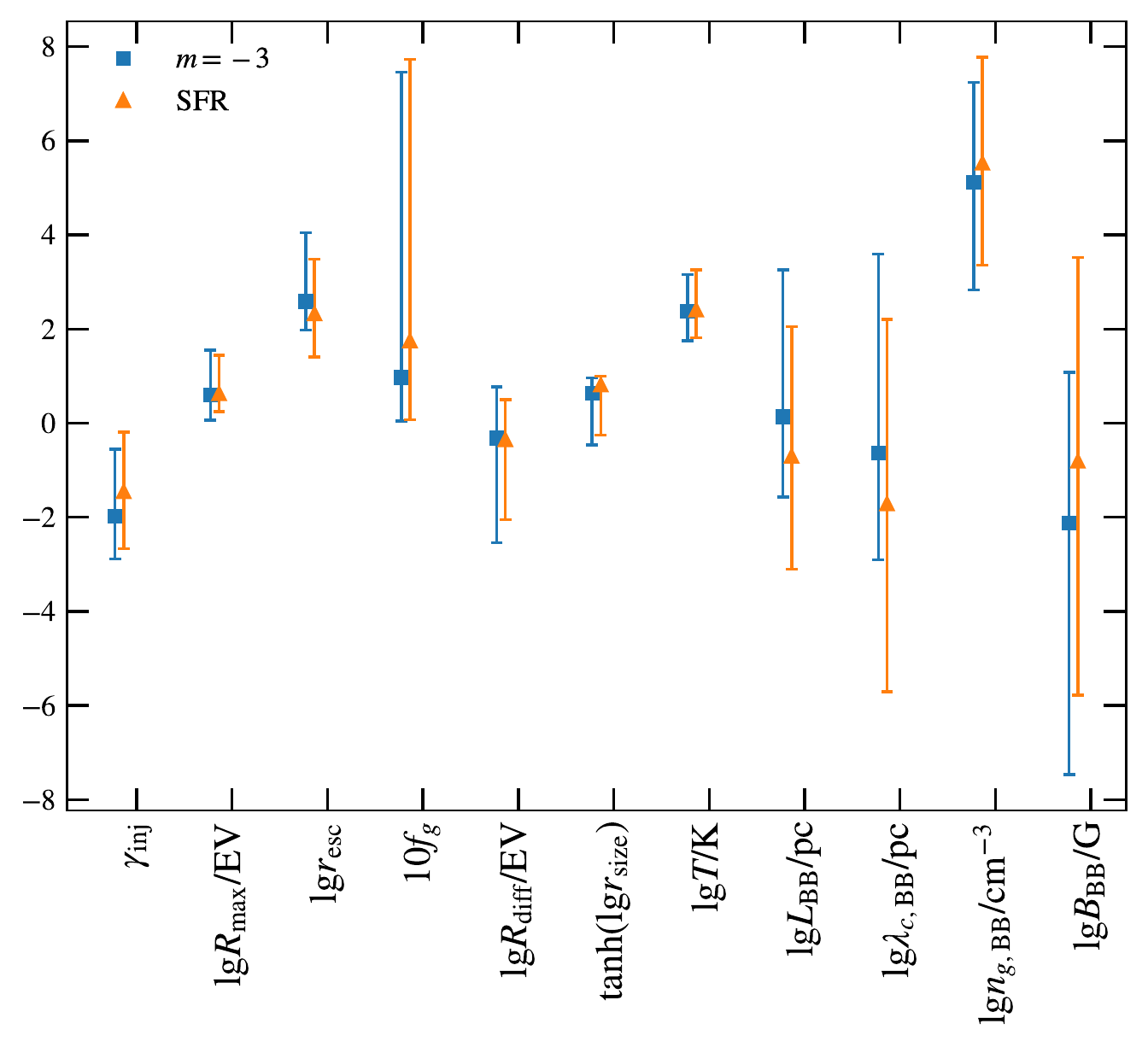}
    \vspace{-0.3in}
	\caption{Same as Fig.~\ref{fig:constraints_comparison} comparing a negative $m=-3$ source (blue) and SFR (orange) evolutions, fitting CR data alone while rejecting models violating EHE neutrino bounds (analagous to the fiducial case above) using the \textsc{Sibyll2.3c} HIM.}
	\label{fig:constraints_comparison_evo}
\end{figure}
\begin{figure*}[htpb!]
	\centering
	\begin{minipage}{0.49\linewidth}
	  \centering
      \subfloat[\label{fig:BL_constraints_negEvo_sibyll}]{\includegraphics[width=\textwidth]{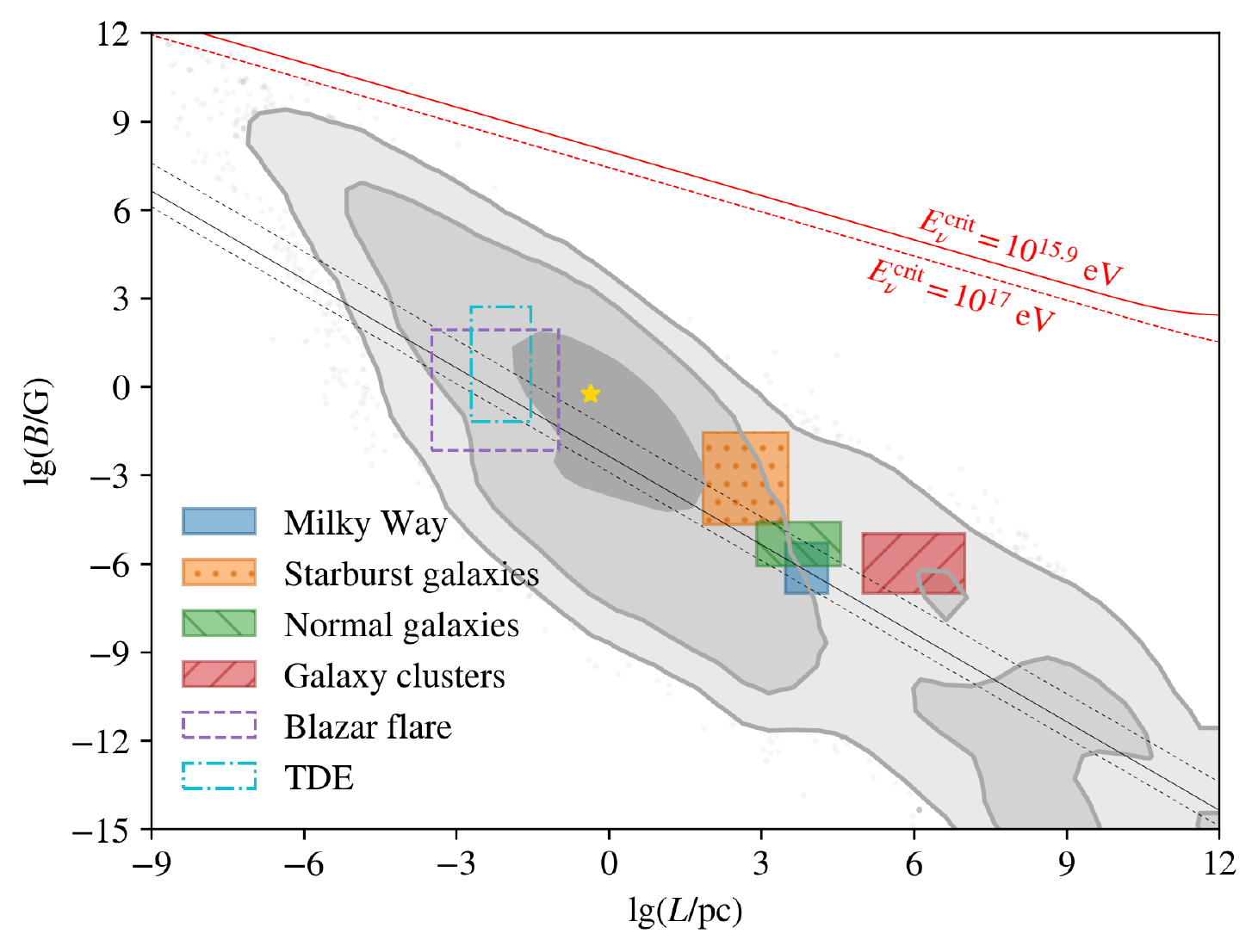}}
    \end{minipage}
    \begin{minipage}{0.49\linewidth}
	  \centering
      \subfloat[\label{fig:BL_LnGas_constraints_negEvo_sibyll}]{\includegraphics[width=\textwidth]{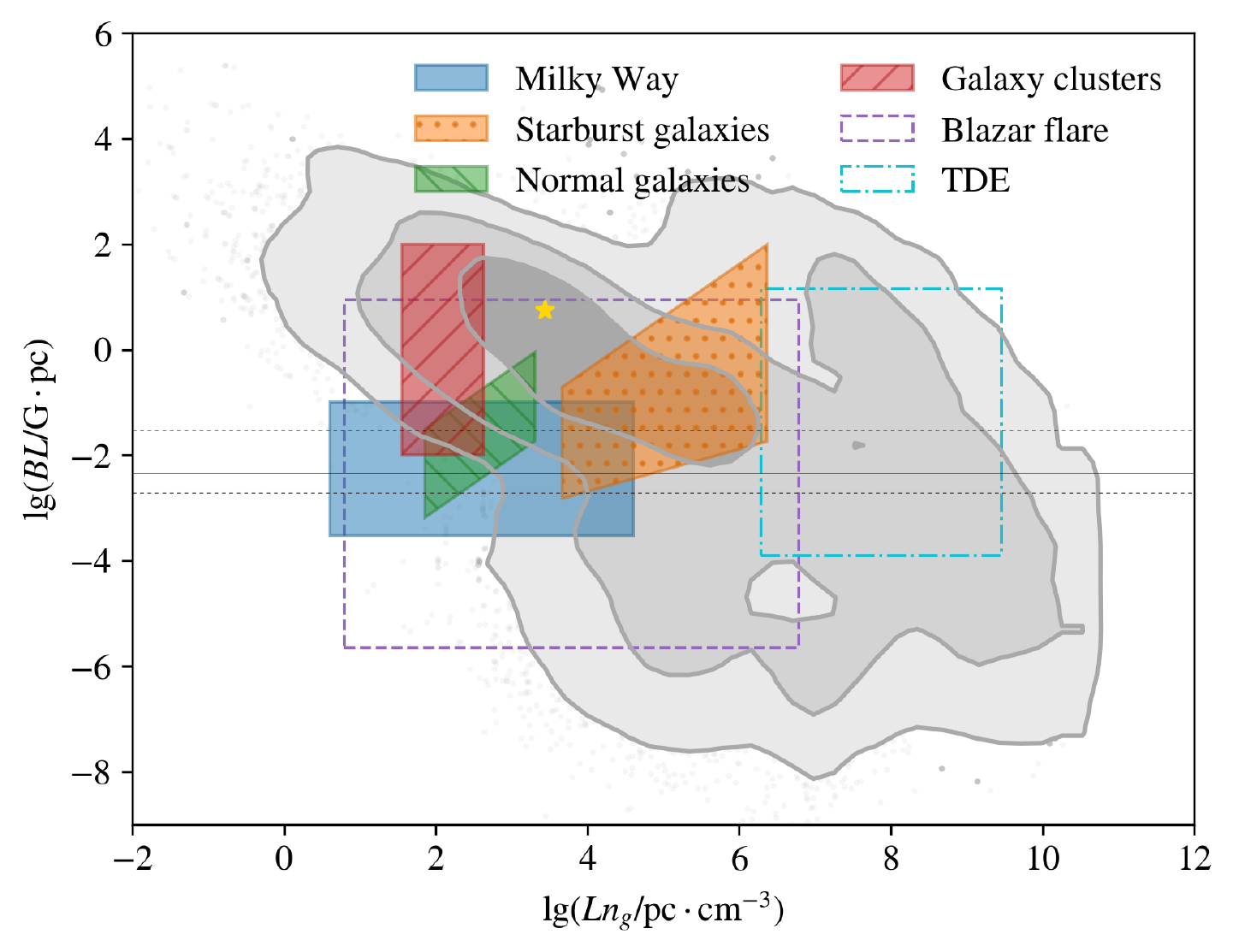}}
    \end{minipage}
	\caption{{\bf Left:} Same as Fig.~\ref{fig:BL_constraints_CR_sibyll} assuming a negative $m=-3$ source evolution. {\bf Right:} Same as Fig.~\ref{fig:BL_LnGas_constraints_CR_sibyll} assuming a negative $m=-3$ source evolution.}
	\label{fig:srcs_constraints_negEvo}
\end{figure*}

Figures~\ref{fig:constraints_comparison}-\ref{fig:astroEnv_CR_epos} and Tables~\ref{tab:parConstraintsNoNus}-\ref{tab:parConstraintsAstroNu} assume a star-formation rate source evolution [SFR,~\citep{Robertson+15}]. For many UHECR source candidates a SFR evolution is not an adequate approximation to their observed evolution. To understand the degree to which our conclusions are sensitive to the assumed source evolution we performed an additional MCMC assuming a source evolution whose CR power density relative to today is given by

\begin{align}
    \xi(z) = 
    \begin{cases} 
      (1+z)^{-3} & z < 2 \\
      (1+z)^{-3} e^{-(z-2)} & z \geq 2
    \end{cases}~,
\end{align}

\noindent
where we have fitted UHECR data alone, rejecting models which violate the IceCube neutrino bounds at the $99\%$~CL [i.e. analogous to our fiducial case above]. Figs.~\ref{fig:constraints_comparison_evo} and \ref{fig:srcs_constraints_negEvo} show how our results change under the assumption of this source evolution. While in detail the results have some differences compared to the SFR case, these results do not change our conclusions. Therefore, we find that the results we present here have very little dependence on the assumed source evolution, for a realistic source evolution.

\section{Viability of starburst galaxies (SBG\texorpdfstring{\MakeLowercase{s}})}\label{app:SBGs}

\par
Understanding whether our analysis favors SBGs as a viable source class is strongly dependent on the grey-body scaling factor, $n_0$, for such systems. To estimate the grey-body factor we considered two model SBGs: 1) M82, representing typical SBGs; and 2) Arp220, representing extremal SBGs. We then fit the peak of their spectral energy distributions (SEDs) with several functional forms (described in Appendix A of~(UFA15): a black-body (BB) spectrum, a modified black-body (MBB) spectrum, and a broken power-law (BPL) spectrum each with an additional parameter controlling their normalization. After fitting for their temperature (or peak energy in the BPL case) and normalization, we were able to extract their grey-body factor as $n_0 \equiv n_\gamma / I_\mathrm{BB}(T)$, where $n_\gamma$ is the integral photon density of the fit and $I_\mathrm{BB}(T)$ is the integral photon density for a pure black-body spectrum of equivalent black-body temperature $T$, as described in~UFA15.

\begin{figure*}[htpb!]
	\centering
	\begin{minipage}{0.49\linewidth}
	  \centering
      \subfloat[\label{fig:M82fit}]{\includegraphics[width=\textwidth]{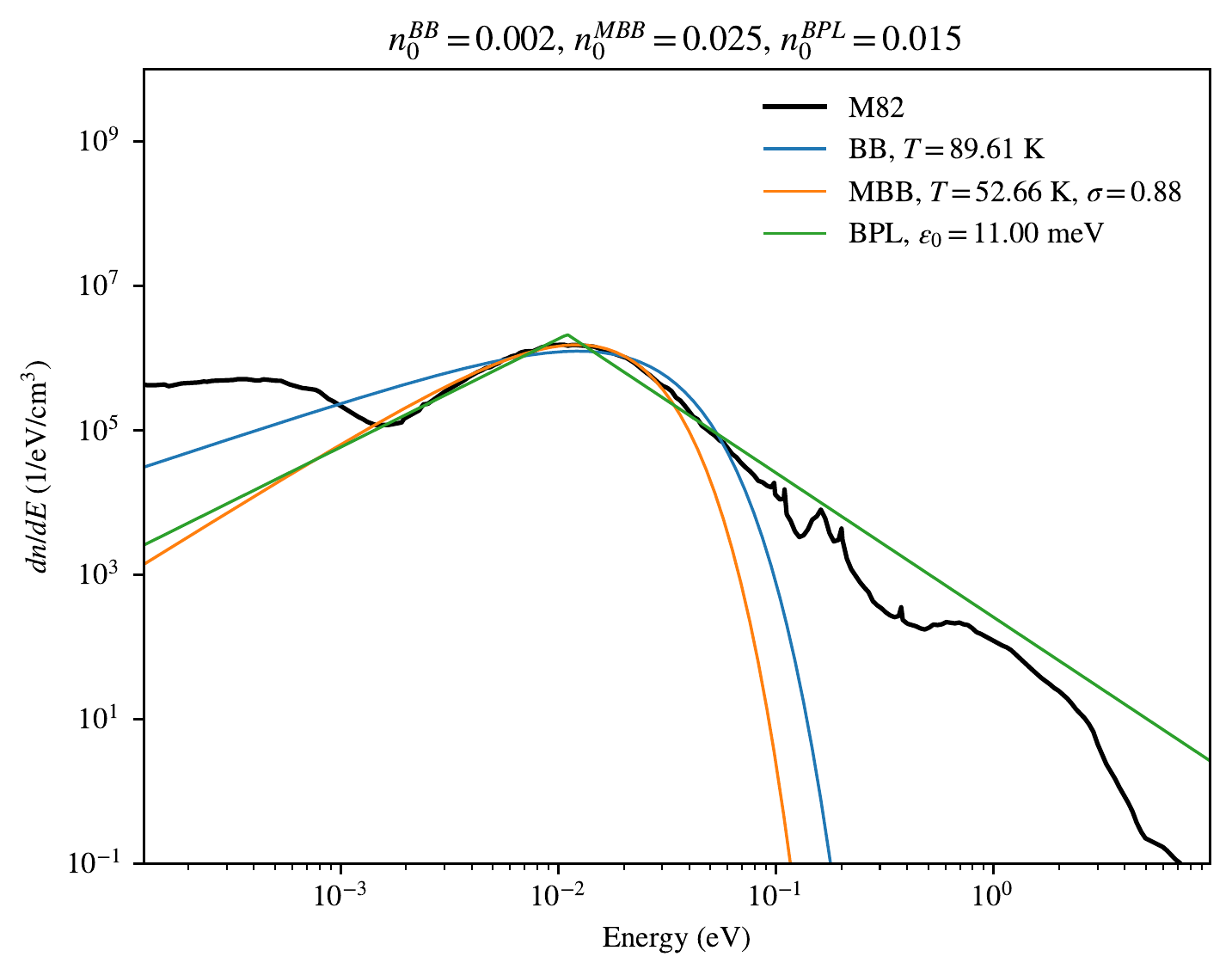}}
    \end{minipage}
    \begin{minipage}{0.49\linewidth}
	  \centering
      \subfloat[\label{fig:Arp220}]{\includegraphics[width=\textwidth]{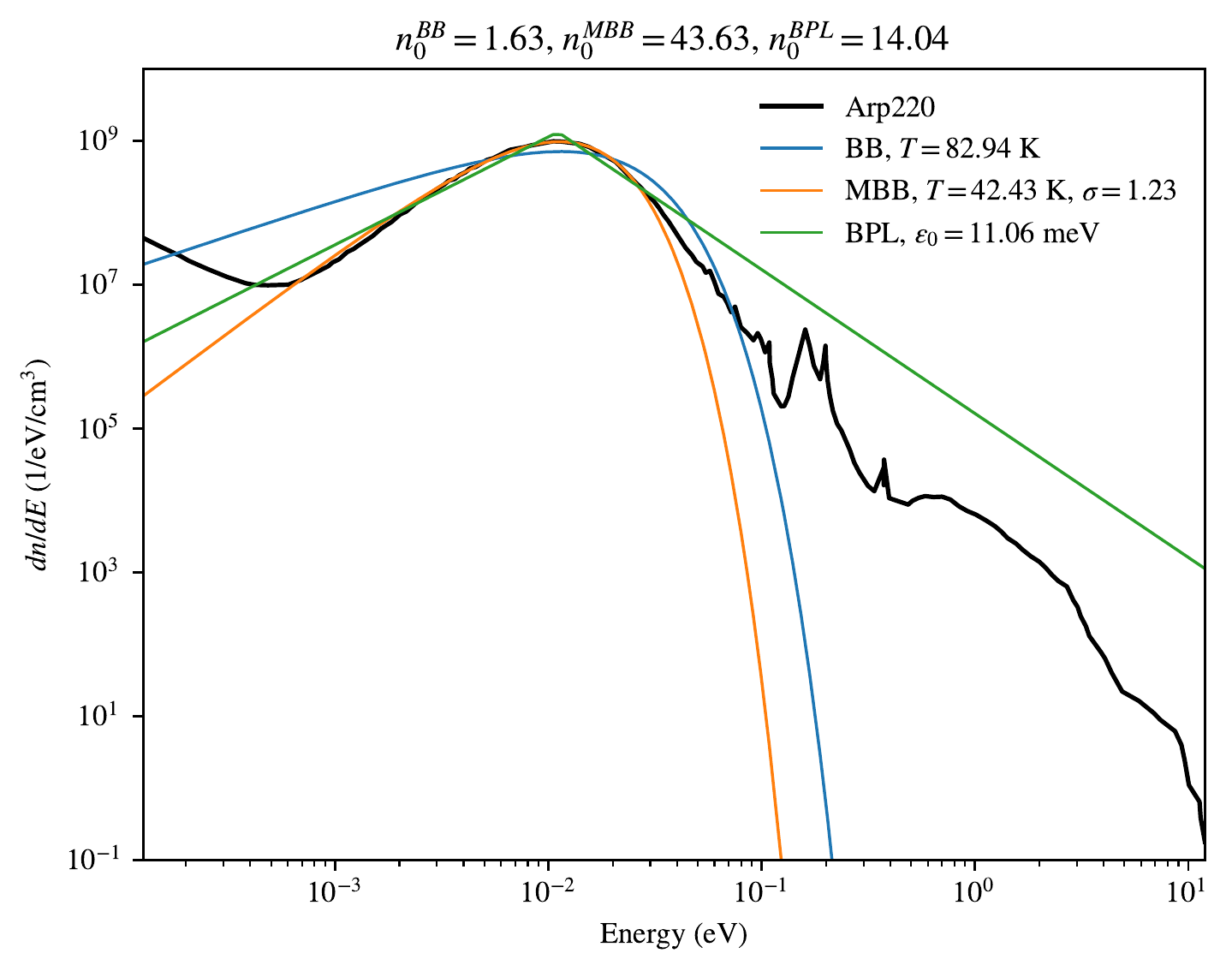}}
    \end{minipage}
	\caption{Best-fits of black-body (BB), modified black-body (MBB), and broken power-law (BPL) spectra to the peak of the SEDs for M82 (left) and Arp220 (right). The best-fit grey-body factors resulting from each of these fits are listed above the figure and provide a rough estimate of the true grey-body factor for each system.}
	\label{fig:SBGfits}
\end{figure*}

\par
For both M82 and Arp220 the SEDs were taken from~\citet{Lacki:2010ue}. Our best-fits are shown in Fig.~\ref{fig:SBGfits}. Fits to both of these systems show that the black-body temperature is fairly consistent at $\sim 80$~K. This temperature is compatible with the lower end of the central $68\%$ of the posterior distribution (see Fig.~\ref{fig:constraints_comparison}). 

\par
The most striking difference, for our purposes, between these two systems is their relative normalization. As the best-fit grey-body factors show, extremal SBGs like Arp220 are much more luminous than typical SBGs like M82. From their SEDs, we infer Arp220 has a grey-body factor $n_0 \sim O(1-10)$, while M82 has a grey-body factor $n_0 \sim O(10^{-3}-10^{-2})$. Translating the posterior distribution in Fig.~\ref{fig:BL_constraints} according to $L=L_\mathrm{BB}/n_0$, $B=B_\mathrm{BB}n_0$, we see that the conditions in M82 are consistent with our analysis of the UHECR data, Arp220 is significantly disfavored.

\bibliography{MF22}{}
\bibliographystyle{aasjournal}

\end{document}